\documentclass[journal=jacsat,manuscript=article]{achemso}
\usepackage[utf8]{inputenc}
\usepackage[T1]{fontenc}

\usepackage[version=3]{mhchem} 
\usepackage{soul}
\usepackage{bm}
\usepackage{threeparttable}
\usepackage{color}
\usepackage{graphicx}


\author{Apurba Nandi}
\email{apurba.nandi@emory.edu}
\affiliation{Department of Chemistry and Cherry L. Emerson Center for Scientific Computation, Emory University, Atlanta, Georgia 30322, U.S.A.}
\author{Riccardo Conte}
\email{riccardo.conte1@unimi.it}
\affiliation{Dipartimento di Chimica, Universit\`{a} Degli Studi di Milano, via Golgi 19, 20133 Milano, Italy}
\author{Chen Qu}
\affiliation{Independent researcher, Toronto, Canada}
\author{Paul L. Houston}
\affiliation{Department of Chemistry and Chemical Biology, Cornell University, Ithaca, New York
14853, U.S.A. and Department of Chemistry and Biochemistry, Georgia Institute of
Technology, Atlanta, Georgia 30332, U.S.A}
\author{Qi Yu}
\affiliation{Department of Chemistry, Yale University, New Haven, Connecticut 06520, U.S.A.}
\author{Joel M. Bowman}
\email{jmbowma@emory.edu}
\affiliation{Department of Chemistry and Cherry L. Emerson Center for Scientific Computation, Emory University, Atlanta, Georgia 30322, U.S.A.}

\title{Quantum calculations on a new CCSD(T) machine-learned PES reveal the leaky nature of gas-phase \textit{trans} and \textit{gauche} ethanol conformers}
\date{\today}

\begin{document}
\newpage

\begin{abstract}
Ethanol is a molecule of fundamental interest in combustion, astrochemistry, and condensed phase as a solvent. It is characterized by two methyl rotors and \textit{trans (anti)} and \textit{gauche} conformers, which are known to be very close in energy. Here we show that based on rigorous quantum calculations of the vibrational zero-point state, using a new ab initio potential energy surface (PES), the ground state resembles the \textit{trans} conformer but substantial delocalization to the \textit{gauche} conformer is present. This explains experimental issues about identification and isolation of the two conformers. This ``leak'' effect is partially quenched when deuterating the OH group, which further demonstrates the need for a quantum mechanical approach. Diffusion Monte Carlo (DMC) and full-dimensional semiclassical dynamics calculations are employed. The new PES is obtained by means of a $\Delta$-Machine learning approach starting from a pre-existing low level (LL) density functional theory (DFT) surface.  This surface is brought to the CCSD(T) level of theory using a relatively small number of \textit{ab initio} CCSD(T) energies. Agreement between the corrected PES and direct \textit{ab initio} results for standard tests is excellent. One- and two-dimensional discrete variable representation calculations focusing on the \textit{trans}-\textit{gauche} torsional motion are also reported, in reasonable agreement with experiment.
\end{abstract}

\flushbottom
\maketitle

\thispagestyle{empty}
\newpage
\section{Introduction}

Ethanol is one of the most important organic molecules with many applications in industrial products,
chemicals, and solvents. It is also the leading biofuel in the transportation sector, where it is mainly used in a form of reformulated gasoline\cite{combust_1, combust_2} and studied from scientific, industrial, and environmental perspectives for its role in internal combustion engines.

Ethanol exists as a mixture of \textit{trans} (or \textit{anti}) and \textit{gauche} ($+/-$) conformers in both solid, liquid, and gaseous state.\cite{Jonsson1976,Richard1980,Luo2015} Therefore, the energy difference between the \textit{trans} and \textit{gauche} conformers is expected to be very small. This is corroborated by the data extracted upon fitting models to spectroscopic experiments in the microwave and far-infrared portion of the electromagnetic spectrum, which estimate the energy gap at 0.12 kcal/mol or 41 cm$^{-1}$ in favor of the more stable \textit{trans} conformer.\cite{Richard1980,Durig1990} Therefore, there is an anticipated preponderance of the \textit{gauche} form at room temperature (62\%) because of its two-fold degeneracy ($+/-$). Furthermore, ethanol has two isomerization saddle points and a three-fold methyl torsional potential, which makes its potential surface much more complex. 

Reports on ethanol in the literature have been often accompanied by several experimental studies of its isomers. In 1980, Quade and co-workers reported microwave torsional-rotational spectra of \textit{gauche}  ethanol\cite{Richard1980} and later Durig and Larsen presented a detailed examination of the torsional modes.\cite{Durig1990} Rotational isomerization of ethanol in nitrogen and argon matrices has been recorded under various conditions of temperature and irradiation in the OH and CO stretches by Coussan $et$ $al$.\cite{RN3987} In 2013, comparative analysis of low-temperature FTIR absorption spectra were reported for ethanol isolated in an argon matrix by Balevicius and co-workers.\cite{RN3983} It was observed that in an argon matrix ethanol is predominantly in the $trans$ configuration, although the most intense absorption lines of the
\textit{gauche} conformer were still observed in the spectra of the samples. Recently, the $trans$-$gauche$ conformational distribution of ethanol has been investigated using the O-H and symmetric C-C-O stretching infrared spectra in argon and nitrogen matrix.\cite{RN3985} It was found that the $trans$ conformer is more populated in nitrogen mixture whereas the $gauche$ conformer is more populated in the argon mixture. After thermal cyclisation in the matrix, the $trans$ conformer isomerises to the $gauche$
conformer in a nitrogen matrix but the reverse happens in an argon matrix. Finally, Pearson \textit{et al.} (PBD) also reported a comprehensive analysis of the threefold asymmetric rotational–torsional spectrum of ethanol in the torsional ground state of the OH internal rotation.\cite{Pearson2008} Zheng et al. considered the partition functions of rotors in ethanol and performed helpful calculations on the energy levels.\cite{Zheng2011}

Ethanol has also been investigated extensively using electronic structure calculations to understand its energetics and complex potential energy surface (PES). In 2004, calculations have been performed at the MP2/aug-cc-pVTZ and CCSD(T)/aug-cc-pVTZ levels of theory by Dyczmons.\cite{Volker2004} It is reported that the $trans$ isomer is 0.52 kJ mol$^{-1}$ or 44 cm$^{-1}$ more stable than the $gauche$ isomer and the energy barrier for the torsional motion of the OH group for $trans$ to $gauche$ isomerisation is 3.9 kJ mol$^{-1}$ or 326 cm$^{-1}$. Recently, a high level calculation has been performed at the CCSD(T)/aug-cc-pVQZ level of theory by Kirschner and co-workers.\cite{Kirschner2018} It was found that the $trans$ isomer is more stable by 0.53 kJ mol$^{-1}$ or 44 cm$^{-1}$ compared to $gauche$ isomer. Thus, it is concluded that the $trans$ conformer is more stable in the gas phase compared to the $gauche$ conformer. Remarkably, a thorough investigation on conformational analysis by systematically improving
the basis set and the level of electron correlation of ethanol has been reported by Kahn and Bruice in 2005.\cite{Kalju2005} Their best estimate of the $trans$-$gauche$ energy gap is 0.134 kcal mol$^{-1}$ or 47 cm$^{-1}$ and the energies of the two isomerization TSs (\textit{eclipsed} and \textit{syn}) are 1.08 kcal mol$^{-1}$ or 378 cm$^{-1}$ and 1.20 kcal mol$^{-1}$ or 420 cm$^{-1}$, respectively, relative to the $trans$ minima. They came to the common conclusion that the $trans$ conformer is more stable in the gas phase compared to the $gauche$ conformer.  Very recently, in 2022, Grimme and co-workers reported combined implicit and explicit solvation protocols for the quantum simulation of ethanol conformers in the gas phase, liquid phase and in \ce{CCl4} solutions. The implicit treatment of solvation effects suggested that the ratio of the \textit{trans} and \textit{gauche} conformers of ethanol increases only slightly when going from gas phase to a \ce{CCl4} solution, and to neat liquid.\cite{grimme22}

However, we note that both experiments and theoretical calculations may not have been conclusive in describing the \textit{trans}-\textit{gauche} dichotomy of gas-phase ethanol.
On the one hand, the experiments referenced above appear to deal with a mixture of the two conformers and to be even affected by experimental conditions. For instance, in Ref.\citenum{RN3987} it is shown that infrared experiments performed in the 8-30K temperature range point at temperature-dependent \textit{trans}-\textit{gauche} isomerism when a nitrogen matrix is employed, while the temperature dependence vanishes and evidence of \textit{trans} isomer only is found when an argon matrix is used. Other experiments found a mixture of the two conformers also in argon matrix, but with abundance conclusions at odds and an inter-conversion rate dependent on temperature and matrix type.
On the other hand, accurate but static theoretical calculations have been performed only at the level of electronic structure, while quantum nuclear effects have not been taken into consideration or have been estimated just with basic and inaccurate harmonic approaches.

The main goal of this study is to investigate the energetics of ethanol and its challenging conformational properties including quantum nuclear effects. This is obtained by means of high-level DMC and semiclassical calculations able to describe nuclear quantum effects performed on a new ``gold standard'' \textit{ab initio} CCSD(T) PES, which we have constructed for this investigation.

Developing high-dimensional, \textit{ab initio}-based PESs remains an active area of theoretical and
computational research. Significant progress has been made in the development of machine learning (ML) approaches to generate PESs for systems with more than five atoms, based on fitting thousands of CCSD(T) energies.\cite{bowman11, ARPC2018, Fu18, guo20} Examples of potentials for 6 and 7-atom chemical reactions which are fits to tens of thousands or even hundred thousand CCSD(T) energies have been reported.\cite{Furoam20, HCH3OH} However, there is a bottleneck for developing the PES at high level theory with the increase of molecular size. Due to the steep scaling of the ``gold standard'' CCSD(T) theory ($\sim N^7$, $N$ being the number of basis functions), it is computationally demanding to fit PESs for systems with a larger and larger number of atoms.

The increasing dimensionality of the PES with the increase of number of atoms requires a large number of training datasets to fit the PES. Thus, the use of lower-level methods such as density functional theory (DFT) and second-order M{\o}ller-Plesset perturbation (MP2) theory is understandable, but probably not accurate enough for precise investigations like the one here targeted. To circumvent this bottleneck, researchers are applying ML approaches to bring a PES based on a low-level of electronic structure theory (DFT or MP2) to a higher level (CCSD(T)) one. One way to achieve this is by means of the $\Delta$-machine learning ($\Delta$-ML) approach, in which a correction is made to a property dataset obtained using an efficient, low-level \textit{ab initio} theory such as DFT or MP2.\cite{Lilienfeld15, Tkatch19, Tkatch2018, Stohr2020, meuwly20, deltaML2021}

We apply the $\Delta$-machine learning approach to take a DFT PES of ethanol to CCSD(T) level upon generation of a manageable subset of \textit{ab initio} CCSD(T) energy points. The new full-dimensional PES is tested against the usual fidelity tests and then employed for the challenging DMC and SC simulations and for an investigation of the wavefunctions of the \ce{-CH3} and \ce{OH} motions, for which Quade et al. have suggested a geared motion by analyzing microwave spectra.\cite{Quade1998a,Quade1998b} 

The paper is organized as follows: In the next Section, we briefly summarize the theory of the $\Delta$-ML approach for PES construction, and diffusion Monte Carlo and adiabatically-switched semiclassical initial value representation for zero-point energy calculations.  Then, we present results with a discussion followed by a summary and conclusions.

\section{Theory and Computational Details}

\subsection{$\Delta$ Machine Learning for PES construction}
The theory underneath the $\Delta$-ML approach is very simple and can be presented in a simple equation
\begin{equation} 
\label{eq:1}
    V_{LL{\rightarrow}CC}=V_{LL}+\Delta{V_{CC-LL}},
\end{equation}
where $V_{LL{\rightarrow}CC}$ is the corrected PES,  $V_{LL}$ is a PES fit to low-level DFT electronic data, and $\Delta{V_{CC-LL}}$ is the correction PES based on high-level coupled cluster energies. It is noted that the difference between CCSD(T) and DFT energies, $\Delta{V_{CC-LL}}$, is not as strongly varying as $V_{LL}$ with respect to the nuclear configurations and therefore just a small number of high-level electronic energies are adequate to fit the correction PES. In the present application to ethanol, we computed a total of 2319 CCSD(T)-F12a/aug-cc-pVDZ electronic energies and performed training on a subset of these data in size of 2069 energies.

Here we employ PIP approach to fit both the $V_{LL}$ and $\Delta{V_{CC-LL}}$ PESs.
The theory of permutationally invariant polynomial is well established and has been presented in several review articles.\cite{Braams2009, Bowman2010, Xie10, bowman11, ARPC2018} In terms of a PIP basis, the potential energy, $V$, can be written in compact form as
\begin{equation}
\label{eq:2}
V(\bm{x})= \sum_{i=1}^{n_p} c_i p_i(\bm{x}),
\end{equation}
where $c_i$ are linear coefficients, $p_i$ are PIPs, $n_p$ is the total number of polynomials for a given maximum polynomial order and $\bm{x}$ are Morse variables.  For example, $x_{\alpha \beta}$ is given by $\exp(-r_{\alpha \beta}/\lambda)$, where $r_{\alpha \beta}$ is the internuclear distance between atoms $\alpha$ and $\beta$. The range (hyper)parameter, $\lambda$, was chosen to be 2 bohr. The linear coefficients are obtained using standard least squares methods for a large data sets of electronic energies (and for large molecules' gradients as well) at scattered geometries.

In order to develop a corrected PES, we need to generate a dataset of high and low-level energies for training and testing. In this study, we need both DFT and CCSD(T) datasets. Training is done for the correction PES $\Delta{V_{CC-LL}}$, and testing is done for the corrected $V_{LL{\rightarrow}CC}$. Do note that this two-step ``training and testing'' is on different datasets.

\begin{figure}[htbp!]
    \includegraphics[width=0.8\columnwidth]{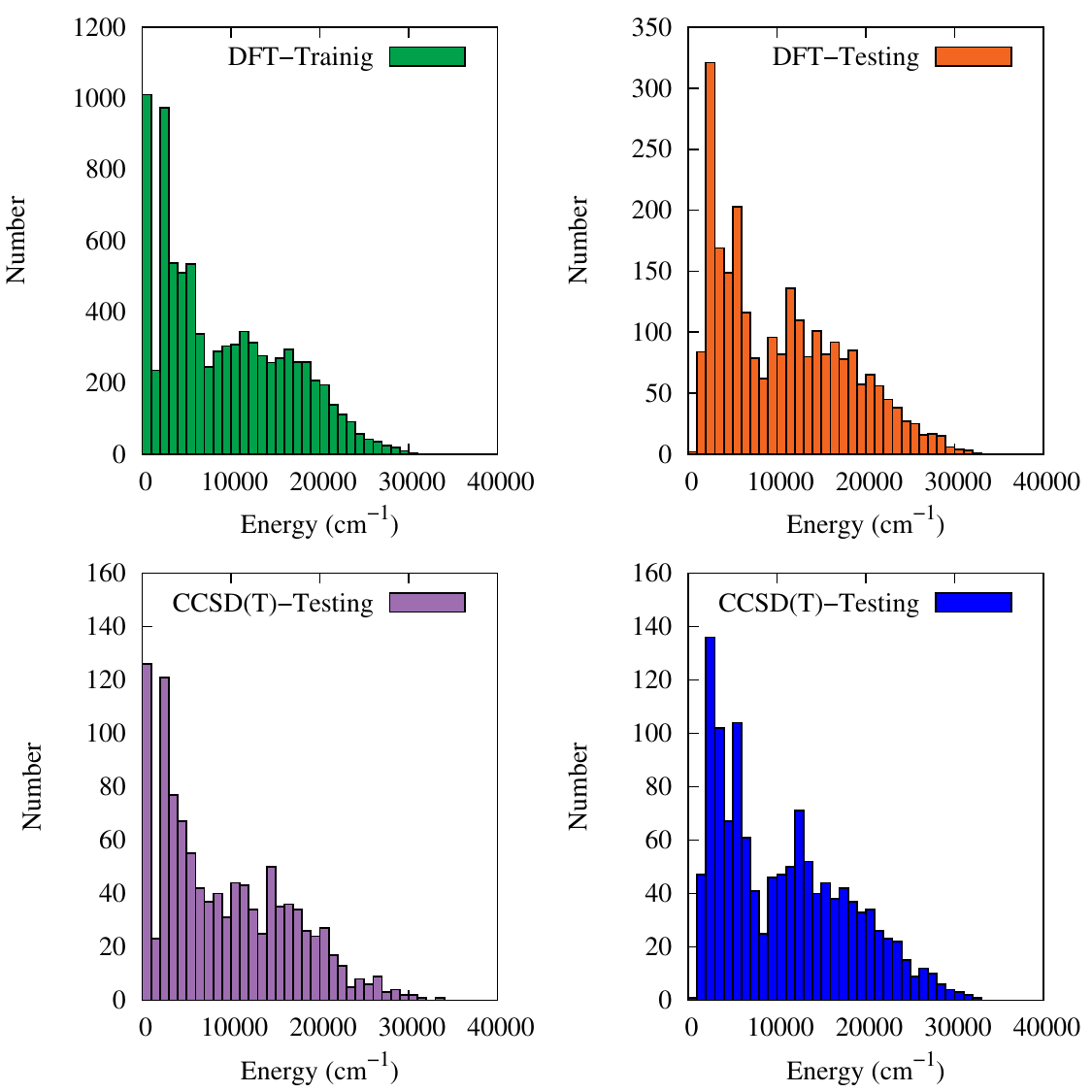}
    \caption{Distributions of DFT and CCSD(T) electronic energies (cm$^{-1}$) of both training and test datasets relative to their respective minimum value.}
        \label{fig:DFT_CCSDT_Dist}
\end{figure}

Here we take the DFT dataset from our recently reported ``MDQM21'' dataset\cite{Bowman_reverse2022} where a total of 11000 energies and their corresponding gradients were generated from \textit{ab initio} molecular dynamics (AIMD) simulations at B3LYP/6-311+G(d,p) level of theory. The DFT PES (V$_{LL}$) was a fit using 8 500 DFT data, which span the energy range of 0–35 000 cm$^{-1}$. Here, we generate a sparse dataset that contains CCSD(T)-F12a/aVDZ energies at 2319 configurations, taken from the ``MDQM21'' dataset. This 2319-geometry dataset is partitioned into a training dataset of 2069 geometries and a test dataset of 250 geometries, respectively. Histogram plots of the distribution of DFT and CCSD(T) electronic energies are shown in Figure \ref{fig:DFT_CCSDT_Dist}, where it can be seen that both the DFT and CCSD(T) datasets span a similar energy range. Geometry optimization and normal-mode analysis are performed to examine the fidelity of the V$_{LL \rightarrow CC}$ PES.

\subsection{Diffusion Monte Carlo}
This PES is also applied to compute rigorous quantum zero-point energies (ZPEs) of ethanol and its single deuterated isotopologues using unbiased DMC calculations. The concept behind DMC is to solve the time-dependent Schr{\"o}dinger equation in imaginary time.\cite{Anderson1975, Anderson1976, Schulten} This is done by simulating a random walk of many replicas, also called ``walkers'', of the molecule, using a birth/death processes. At each step, a random displacement in each degree of freedom is assigned to each walker, and this walker may remain alive (and may give birth to a new walker) or be killed by comparing its potential energy, $E_i$, with a reference energy, $E_r$. For the ground state, the probability of birth or death is given as:
\begin{align}
    & P_\text{birth} = \exp \left[ -(E_i - E_r)\Delta \tau\right] - 1 \ (E_i < E_r)\\
    & P_\text{death} = 1 - \exp \left[ -(E_i - E_r)\Delta \tau\right] \ (E_i > E_r),
\end{align}
where $\Delta \tau$ is the step size in imaginary time.
After removing all dead walkers, the reference energy is updated using the equation
\begin{equation}
  E_r(\tau) = \langle V(\tau) \rangle - \alpha \frac{N(\tau) - N(0)}{N(0)},
\end{equation}
where $\tau$ is the imaginary time; $\langle V(\tau) \rangle$ is the average potential over all the walkers that are alive; $N(\tau)$ is the number of live walkers at time $\tau$; $\alpha$ is a parameter that can control the fluctuations in the number of walkers and the reference energy. Finally, the average of the reference energy over the imaginary time gives an estimate of ZPE.

In this study, each DMC trajectory is propagated for 30,000 time steps with step size of $5.0$ a.u.; 20,000 steps are used to equilibrate the walkers, and the reference energies in the remaining 10,000 steps are used to compute the ZPE. For each isomer, 15 DMC simulations (or trajectories) were carried out, and the final ZPE is the average of the 15 simulations. Statistical uncertainty of the zero-point energy is defined as the standard deviation of DMC energies over the total number of simulations. This is written as
\begin{equation}
\Delta E =\sqrt{\frac{1}{15}\sum\limits_{i=1}^{15} (E_i - \bar E)^2},
\end{equation}
where $\bar E$ is the average energy over the 15 simulations. We also perform DMC calculations on three single deuterated isotopologues employing 15 DMC trajectories. For $trans$- and $gauche$-\ce{CH3CH2OH} and  $trans$- and $gauche$-\ce{CH3CH2OD} 40,000 random walkers are used, while for \ce{CH2DCH2OH} and \ce{CH3CHDOH} only 20,000 random walkers are employed.

\subsection{Adiabatically Switched Semiclassical Initial Value Representation}
Another application of the PES that we present concerns the calculation of ethanol (\textit{trans} and \textit{gauche}) ZPEs, and those of its deuterated isotopologues, by means of the adiabatically switched semiclassical initial value representation (AS-SCIVR) technique.\cite{Conte_Ceotto_ASSCIVR_2019,Botti_Conte_ASotf_2022} AS-SCIVR is a recently developed two-step semiclassical approach able to regain quantum effects starting from classical trajectories. Under this aspect it is quite similar to standard semiclassical techniques\cite{miller2001semiclassical,Huber_Heller_SCdynamics_1987} by which it differs in the way the starting conditions of the semiclassical dynamics run are selected. In AS-SCIVR a preliminary adiabatic switching dynamics is performed. This allows to start from harmonic quantization and approximately preserve it after switching on the true system Hamiltonian. The exit molecular geometry and momenta of the adiabatic switching run serve as starting conditions for the subsequent semiclassical dynamics trajectory, and this procedure is applied to a distribution of harmonically quantized starting conditions.

In practice the adiabatic switching Hamiltonian is\cite{Qiyan_Gazdy_AdiabaticSwitching_1988,Saini_Taylor_AdiabaticSwitching_1988,Nagy_Lendvay_AdiabaticSwitching_2017}

\begin{equation}
    H_{\mathrm{as}} = \left[ 1-\lambda (t) \right] H_{\mathrm{harm}} + \lambda (t) H_{\mathrm{anh}},
    \label{eqn:switching_ham}
\end{equation}
where $\lambda (t)$ is the following switching function
\begin{equation}
    \lambda(t) = \frac{t}{T_{\mathrm{AS}}} - \frac{1}{2\pi} \sin \left( \frac{2\pi t}{T_{\mathrm{AS}}} \right),
    \label{eqn:switching}
\end{equation}
$H_{harm}$ is the harmonic Hamiltonian built from the harmonic frequency of vibration calculated by Hessian matrix diagonalization at the equilibrium geometry ${\bf q}_{eq}$, and $H_{anh}$ is the actual molecular vibrational Hamiltonian. In our simulations $T_{AS}$ has been chosen equal to 25000 a.u. (about 0.6 ps) and time steps of 10 a.u. have been employed. 5400 trajectories were evolved starting from harmonic ZPE quantization.

Once the adiabatic switching run is over, the trajectories are evolved according to $H_{anh}$ for another 25000 a.u. with same step size to collect the dynamical data needed for the semiclassical calculation. This relies on Kaledin and Miller's time-averaged version of semiclassical spectroscopy. Therefore, the working formula is 
\begin{equation}
 I_{as} (E) =\left( \frac{1}{2\pi \hbar} \right)^{N_{v}} \sum_{i=1}^{N_{traj}}\frac{1}{2\pi\hbar T} \left\vert \int_{0}^{T} \, dt e^{\frac{i}{\hbar} \left[ S_{t} (\mathbf{p}_{as},\mathbf{q}_{as}) + Et + \phi_{t} (\mathbf{p}_{as},\mathbf{q}_{as}) \right]} \langle \Psi (\mathbf{p}_{eq},\mathbf{q}_{eq}) \vert g (\mathbf{p}^{\prime}_{t},\mathbf{q}^{\prime}_{t}) \rangle \right\vert^{2},
    \label{eqn:astascivr}
\end{equation}
where $I_{as} (E )$ indicates that a vibrational spectral density is calculated as a function of the vibrational energy $E$. $I_{as}$ is peaked at the eigenvalues of the vibrational Hamiltonian, the lowest one being the ZPE. 
Eq.(\ref{eqn:astascivr}) is made of several terms. $N_v$ is the number of vibrational degrees of freedom of the system, i.e. 21 in the case of ethanol.  $T$ is the total evolution time of the dynamics for the semiclassical part of the simulation. As anticipated, we chose $T$ equal to 25000 a.u. with a time step size of 10 a.u. $({\bf p}_{t}^\prime,{\bf q}_{t}^\prime)$ is the instantaneous full-dimensional phase space trajectory started at time 0 from the final phase space condition $({\bf p}_{as},{\bf q}_{as})$ of the adiabatic switching part of the simulation. $S_t$ is the classical action along the semiclassical trajectory, and $\phi_t$ is the phase of the Herman-Kluk pre-exponential factor based on the elements of the stability matrix and defined as 
\begin{equation}
    \phi_{t} = \mathrm{phase} \left[ \sqrt{\left\vert \frac{1}{2} \left( \frac{\partial \mathbf{q}^\prime_{t}}{\partial \mathbf{q}_{as}} + \Gamma^{-1} \frac{\partial \mathbf{p}^\prime_{t}}{\partial \mathbf{p}_{as}} \Gamma -i\hbar \frac{\partial \mathbf{q}^\prime_{t}}{\partial \mathbf{p}_{as}} \Gamma + \frac{i \Gamma^{-1}}{\hbar} \frac{\partial \mathbf{p}^\prime_{t}}{\partial \mathbf{q}_{as}} \right) \right\vert} \right],
    \label{eqn:phase}
\end{equation}
where $\Gamma$ is an $N_v\times N_v$ matrix usually chosen to be diagonal with elements numerically equal to the harmonic frequencies. We note that evolution in time of $\phi_t$ requires calculation of the Hessian matrix, which represents the bottleneck of the AS-SCIVR approach and semiclassical methods broadly speaking. Based on Liouville's theorem, the stability (or monodromy) matrix has the property to have its determinant equal to 1 along the entire trajectory. However, classical chaotic dynamics can lead to numerical inaccuracies in the propagation, so, following a common procedure in semiclassical calculations, we have rejected the trajectories based on a 1\% tolerance threshold on the monodromy matrix determinant value.  Finally, the working formula is completed by a quantum mechanical overlap between a quantum reference state $|\Psi\rangle$ and a coherent state $|g\rangle$ characterized by the following representation in configuration space 
\begin{equation}
    \langle\mathbf{q}|g(\mathbf{p}^{\prime}_{t},\mathbf{q}^{\prime}_{t})\rangle = \left( \frac{\det (\Gamma)}{\pi^{N_{\nu}}} \right) \exp \left\lbrace - (\mathbf{q}-\mathbf{q}^\prime_{t})^{T} \frac{\Gamma}{2} (\mathbf{q}-\mathbf{q}^\prime_{t}) + \frac{i}{\hbar}\mathbf{p}^{\prime T}_{t}(\mathbf{q}-\mathbf{q}^\prime_{t}) \right\rbrace.
    \label{eqn:coherent_superposition}
\end{equation}
The reference state $|\Psi\rangle$ is usually chosen to be itself a coherent state. In Eq. (\ref{eqn:astascivr}) $|\Psi\rangle$ is written as $|\Psi (\mathbf{p}_{eq},\mathbf{q}_{eq})\rangle$, where ${\bf p}_{eq}$ stands for the linear momenta obtained in harmonic approximation setting the geometry at the equilibrium one (${\bf q}_{eq}$).

\section{Results and Discussion}
\subsection{The starting low level PES (V$_{LL}$)}

The low level PES, V$_{LL}$ is developed using the efficient B3LYP/6-311+G(d,p) level of theory. 
For the fit, we use maximum polynomial order of 4 with permutationally symmetry 321111, which leads to a total of 14752 PIPs in the fitting basis set. These are used to fit a dataset of 8500 energies and their corresponding gradients. The fitting RMS errors for energies and gradients are 40 cm$^{-1}$ and 73 cm$^{-1}$ bohr$^{-1}$, respectively. Testing is done on 2500 geometries. The testing RMS errors for energies and gradients are 51 cm$^{-1}$ and 106 cm$^{-1}$ bohr$^{-1}$, respectively. This PES is not new and it has already been reported in our recent work.\cite{Bowman_reverse2022}

\subsection{The correction PES ($\Delta$V$_{CC-LL}$)}

A dataset of 2319 geometries are sparsely selected from the ``MDQM21'' DFT dataset and CCSD(T)-F12a/aug-cc-pVDZ energy computations are performed at those geometries. To develop the correction PES, we train $\Delta$V$_{CC-LL}$ on the difference between the CCSD(T) and DFT absolute energies of 2069 geometries and test the obtained surface on the remaining 250 geometries. A plot of $\Delta$V$_{CC-LL}$ versus the DFT energies is shown in the Supplementary Information (SI) file for both training and test datasets. Note that we reference $\Delta$V$_{CC-LL}$ to the minimum of the difference between the CCSD(T) and DFT energies (roughly 35 732 cm$^{-1}$). As seen, the energy range of $\Delta$V$_{CC-LL}$ is about 1800 cm$^{-1}$, which is much smaller than the DFT energy range relative to the minimum value (roughly 35 000 cm$^{-1}$).

The difference $\Delta$V$_{CC-LL}$ is not as strongly varying as
V$_{LL}$ with respect to the nuclear configuration. Therefore, low-order polynomials will be adequate to fit the correction PES. We use maximum polynomial order of 2 with permutational symmetry 321111 to fit the training dataset which leads to a total of 208 unknown linear coefficients (equivalent to the number of terms in the PIP fitting basis set). These coefficients are determined by solving a linear least-squares problem. The PIP basis to fit this PES is generated using our ``in-house'' MSA software.\cite{NandiQuBowman2019,msachen} The fitting RMS error of this $\Delta$V$_{CC-LL}$ fit is 25 cm$^{-1}$. The fit is tested on the 250 energy differences and the RMS test error in this case is 41 cm$^{-1}$.



\subsection{The New CCSD(T) Ethanol PES (V$_{LL \rightarrow CC}$)}

\begin{figure}[htbp!]
    \includegraphics[width=0.8\columnwidth]{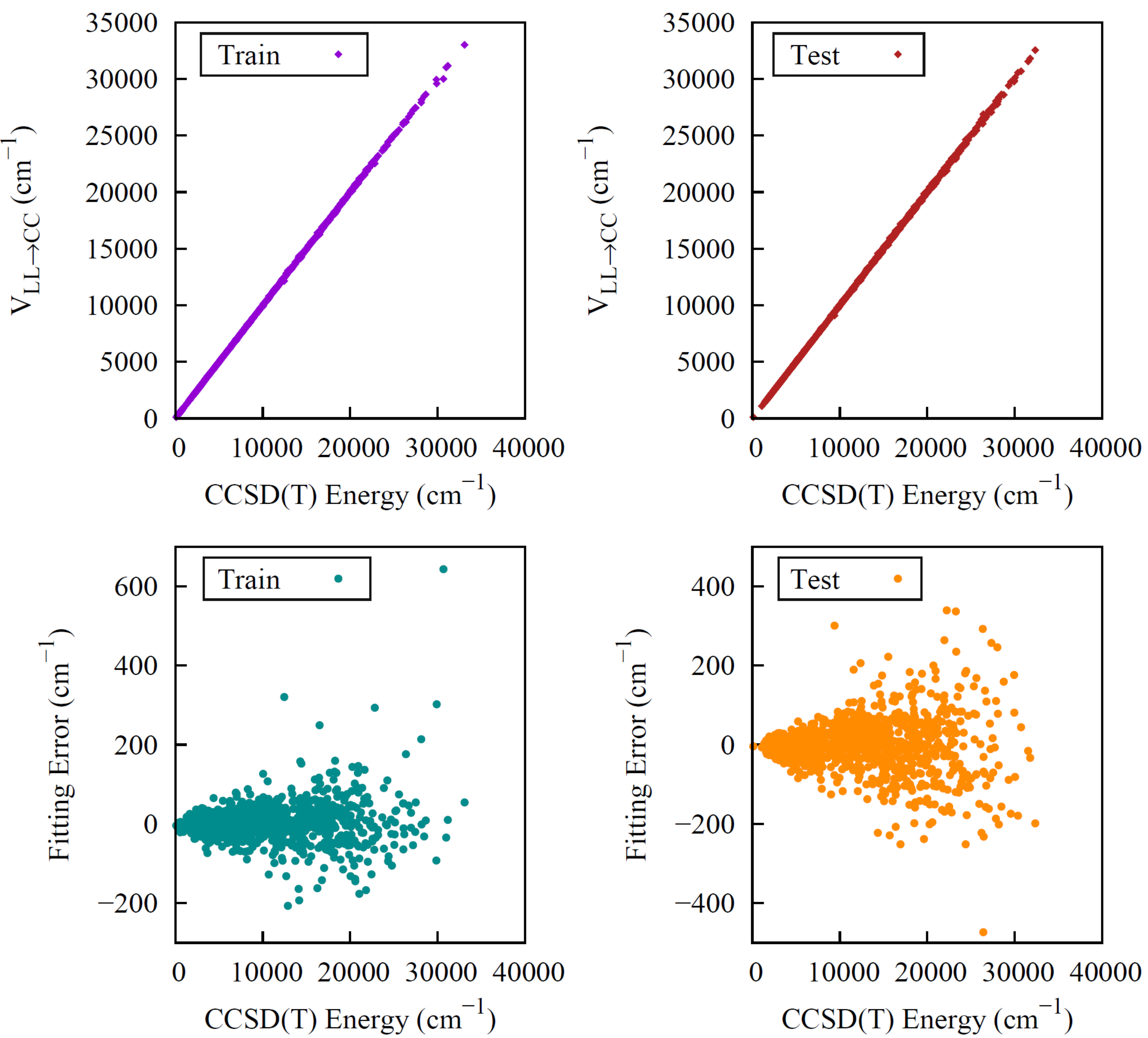}
    \caption{The two upper panels show energies of \ce{CH3CH2OH} from $V_{LL{\rightarrow}CC}$ vs direct CCSD(T) ones for the indicated data sets.  The one labeled ``Train'' corresponds to the configurations used in the training of $\Delta{V_{CC-LL}}$ and the one labeled ``Test'' is just the set of remaining configurations.  Corresponding fitting errors relative to the minimum energy are given in the lower panels.}
        \label{fig:CCSD_fit}
\end{figure}

To obtain the CCSD(T) energies we add the correction $\Delta$V$_{CC-LL}$ to the low-level DFT PES, V$_{LL}$. A plot of V$_{LL \rightarrow CC}$ vs corresponding direct CCSD(T) energies for the training set of 2069 points and the test set of 250 points is shown in Figure \ref{fig:CCSD_fit}. As seen, there is overall excellent precision; however, we see a few larger errors. The RMS differences between the  V$_{LL \rightarrow CC}$ and direct CCSD(T) energies for the training and test datasets are 49 cm$^{-1}$ and 63 cm$^{-1}$, respectively. 

\begin{figure}[htbp!]
    \includegraphics[width=0.6\columnwidth]{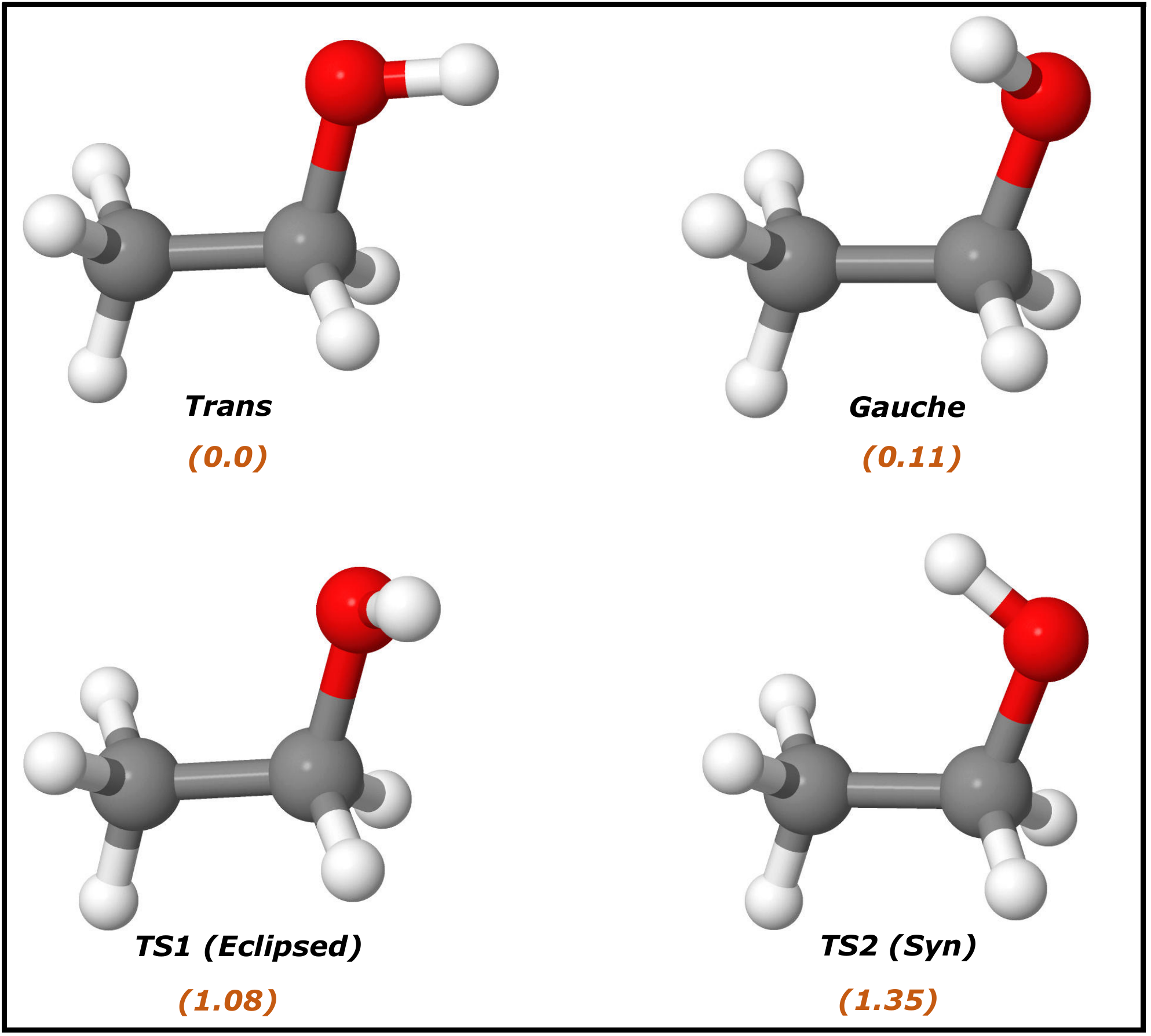}
    \caption{Geometry of $trans$ and $gauche$ conformers of ethanol and their two isomerization TSs and their electronic energies (kcal/mol) relative to the $trans$ minimum from $\Delta$-ML PES.}
        \label{fig:geom}
\end{figure}

To examine this fidelity of the new V$_{LL \rightarrow CC}$ PES, we perform geometry optimization and normal mode frequency calculation of both \textit{trans} and \textit{gauche} isomers and their two isomerization saddle point geometries. They are the \textit{eclipsed} one, in which the hydroxylic hydrogen eclipses with the hydrogen of the adjacent \ce{CH2} group, and the \textit{syn} one, in which the hydroxylic hydrogen is above the methyl group. The structures of these isomers and saddle points are shown in Figure \ref{fig:geom}. We get the PES optimized energies within 5 cm$^{-1}$ of the direct CCSD(T)-F12a calculation and find that the \textit{trans} isomer is lower in energy by 38 cm$^{-1}$.
Next, to examine the vibrational frequency predictions of the
PES, we perform normal mode analyses for both \textit{trans} and \textit{gauche} isomers and their isomerization saddle points. The comparison of harmonic mode frequencies of \textit{trans} and \textit{gauche} ethanol with their corresponding \textit{ab initio} ones are shown in Table \ref{tab:PES_Freq}. The agreement with the direct CCSD(T)-F12a/aug-cc-pVDZ frequencies is overall very good; the maximum error is 21 cm$^{-1}$ for the lowest frequency mode of \textit{trans} conformer, but most of the frequencies are within a few cm$^{-1}$ of the \textit{ab initio} ones and the mean absolute error (MAE) is only 4 cm$^{-1}$. The \textit{gauche} isomer shows even better agreement with the \textit{ab initio} data. The two $trans$ - $gauche$ isomerization saddle point geometries such as \textit{eclipse} and \textit{syn} ones are confirmed by obtaining one imaginary frequency. The normal mode frequencies of this saddle point geometry are given in the Supporting Information file. The barrier height of $trans$ - $gauche$ isomerization with respect to eclipse and syn TSs are found to be 377 cm$^{-1}$ and 472 cm$^{-1}$, respectively, and the corresponding direct \textit{ab initio} values are 389 cm$^{-1}$ and 438 cm$^{-1}$. These are in excellent agreement with the experimental barrier heights of 402 cm$^{-1}$ and 444 cm$^{-1}$.\cite{Durig1990}


   


\begin{table}[htbp!]
\centering
\caption{Comparison of harmonic frequencies (in cm$^{-1}$) between $V_{LL{\rightarrow}CC}$ PES and the corresponding \textit{ab initio} (CCSD(T)-F12a/aug-cc-pVDZ) ones of both \textit{trans} and \textit{gauche} isomers of ethanol.}
\label{tab:PES_Freq}

	\begin{tabular*}{0.9\columnwidth}{@{\extracolsep{\fill}}rrrrrrr}
	\hline
	\hline\noalign{\smallskip}
	& \multicolumn{3}{c}{$trans$-ethanol} & \multicolumn{3}{c}{$gauche$-ethanol} \\
	\noalign{\smallskip} \cline{2-4} \cline{5-7} \noalign{\smallskip}
     Mode & \ $\Delta$-ML PES\  & \ \textit{ab initio}\ & \ Diff.\ & \ $\Delta$-ML PES\  & \ \textit{ab initio}\ & \ Diff.\ \\
	\noalign{\smallskip}\hline\noalign{\smallskip}
       1 &  243 &  222 & -21 &  268 &  258 & -10 \\
       2 &  273 &  274 &   1 &  278 &  271 &  -7 \\
       3 &  417 &  413 &  -4 &  424 &  420 &  -4 \\
       4 &  818 &  813 &  -5 &  804 &  803 &  -1 \\ 
       5 &  909 &  907 &  -2 &  894 &  895 &   1 \\
       6 & 1055 & 1049 &  -6 & 1075 & 1069 &  -6 \\
       7 & 1115 & 1115 &   0 & 1094 & 1096 &   2 \\
       8 & 1181 & 1180 &  -1 & 1144 & 1141 &  -3 \\
       9 & 1284 & 1274 & -10 & 1290 & 1284 &  -6 \\
      10 & 1302 & 1300 &  -2 & 1375 & 1374 &  -1 \\
      11 & 1403 & 1402 &  -1 & 1406 & 1402 &  -4 \\
      12 & 1454 & 1456 &   2 & 1424 & 1426 &   2 \\
      13 & 1488 & 1484 &  -4 & 1490 & 1491 &   1 \\
      14 & 1500 & 1501 &   1 & 1496 & 1497 &   1 \\
      15 & 1530 & 1531 &   1 & 1519 & 1522 &   3 \\
      16 & 2995 & 3001 &   6 & 3007 & 3014 &   7 \\
      17 & 3028 & 3036 &   8 & 3020 & 3028 &   8 \\
      18 & 3036 & 3042 &   6 & 3088 & 3089 &   1 \\
      19 & 3120 & 3122 &   2 & 3108 & 3108 &   0 \\
      20 & 3126 & 3127 &   1 & 3121 & 3123 &   2 \\
      21 & 3862 & 3853 &  -9 & 3845 & 3837 &  -8 \\
    \noalign{\smallskip}\hline
	\hline
	\end{tabular*}

\end{table}

\begin{figure}[htbp!]
    \includegraphics[width=0.6\columnwidth]{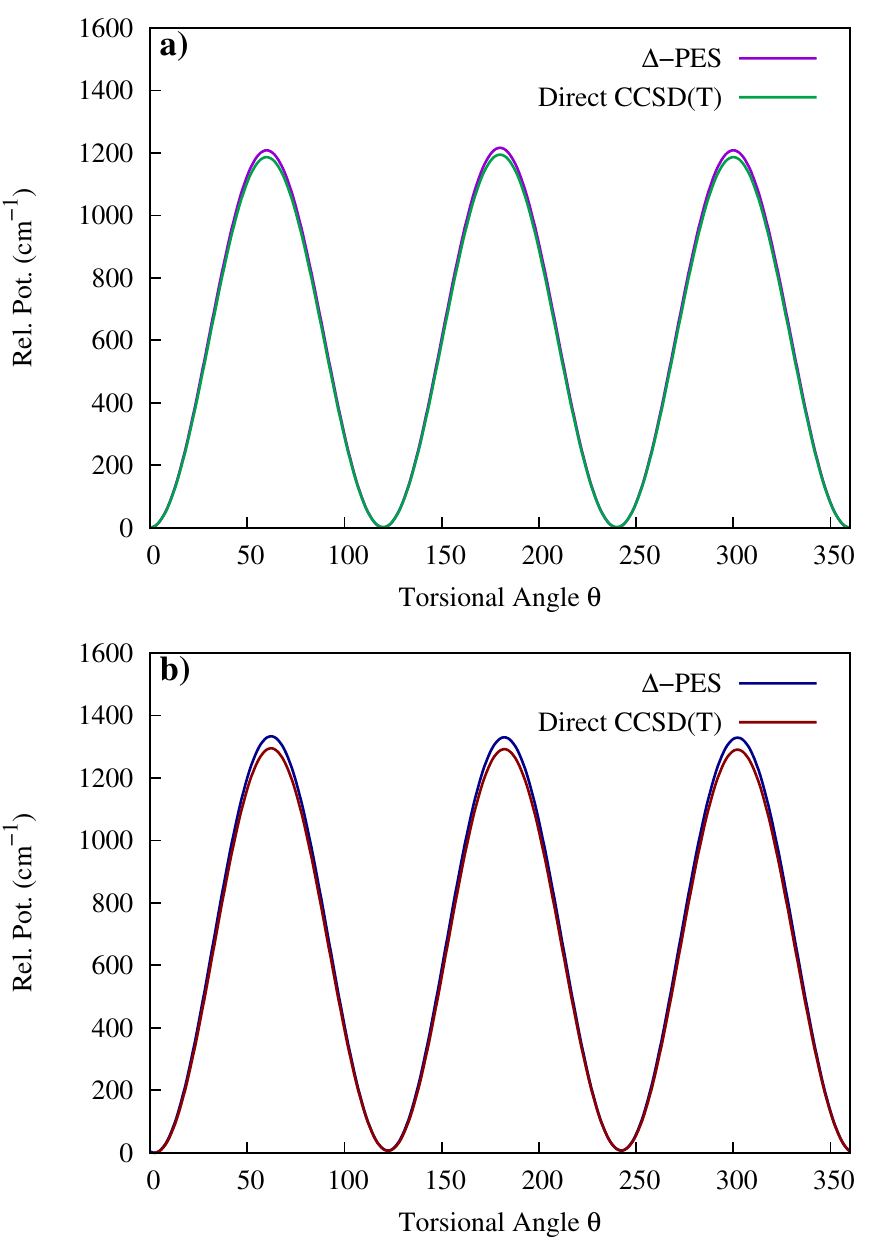}
    \caption{Comparison of torsional potential (not fully relaxed) of the methyl rotor of $trans$ (a) and $gauche$ (b) Ethanol between direct CCSD(T) and $\Delta$-ML PES.}
        \label{fig:torsional}
\end{figure}

Another comparison to the experiment we are able to perform thanks to the new PES concerns the torsional barrier for the methyl rotor. The methyl rotor torsional potentials (not fully relaxed) for both \textit{trans} and \textit{gauche} isomers as a function of the torsional angle are shown in Figure \ref{fig:torsional}. It is seen that results from the PES are very close to the ones obtained from direct \textit{ab initio} calculations at CCSD(T) level by means of a set of single point calculations. We obtain that the methyl torsional barriers for $trans$ and $gauche$ isomers are 1208 cm$^{-1}$ and 1324 cm$^{-1}$, respectively. The methyl torsional barrier heights extrapolated from microwave spectroscopy for the $trans$ and $gauche$ isomers are 1174 cm$^{-1}$ and 1331 cm$^{-1}$.\cite{Richard1980,Quade2000,Pearson1995} A different experimental analysis of the infrared and Raman spectra determined the methyl torsional barriers to be 1185 cm$^{-1}$ and 1251 cm$^{-1}$ for $trans$ and $gauche$, respectively.\cite{Durig1990} To complete our investigation of torsional barriers, in the Supporting Information we report the methyl rotor torsional potential (not fully relaxed) for TS1 and TS2 geometries as a function of the \ce{CH3} torsional angle. We get perfect three-fold symmetry with barrier heights of 1283 and 1404 cm$^{-1}$, respectively. 

This is another proof of the accuracy of the new PES and another evidence of experimental results obtained from ethanol vibrational spectroscopy being not conclusive. So far only electronic energies have been investigated, but we now move to consider nuclear quantum effects.

As a remarkable quantum nuclear application of the PES, we present the results of diffusion Monte Carlo (DMC) calculations of the zero-point energy (ZPE) for both \textit{trans} and \textit{gauche} isomers and singly deuterated isotopologues. In addition to that, it is well known that a DMC calculation is a very challenging test to examine the quality of a PES in extended regions of the configuration space. A common issue in PES fitting is the unphysical behavior in the extrapolated regions where the fitting dataset is lacking data, and this is dramatically manifested by large negative values. These are referred to as ``holes'' in the PES. Generally, we have observed that ``holes'' occur for highly repulsive configurations, i.e., short internuclear distances. Adding some more data in these regions and perform a refit generally eliminates the issue. So, one goal of presenting DMC calculations is also to demonstrate that our PES correctly describes the high energy regions of ethanol and it is therefore suitable for quantum approaches that need to sample these regions.

\begin{table}[htbp!]
\caption{Harmonic, DMC, and SC ZPEs (cm$^{-1}$) of $trans$ and $gauche$  ethanol and singly deuterated isotopologues. The zero of energy is set at the electronic global minimum. Values inside the parentheses represent statistical uncertainties in the DMC results.} 
\label{tab:DMC}

\begin{threeparttable}
	\begin{tabular*}{\columnwidth}{@{\extracolsep{\fill}}llll}
	\hline
	\hline\noalign{\smallskip}
     Molecule & Harmonic ZPE & DMC ZPE & SC ZPE\\
	\noalign{\smallskip}\hline\noalign{\smallskip}
     \ce{CH3CH2OH}(\textit{trans}) & 17568  &  17321 (9) & 17298\\
     \ce{CH3CH2OH}(\textit{gauche}) & 17621 &  17321 (6) & 17317\\
     \ce{CH3CH2OD}(\textit{trans}) & 16842 &   16619 (6) & 16598 \\
     \ce{CH3CH2OD}(\textit{gauche}) & 16894 &  16619 (8) & 16611 \\
     \ce{CH2DCH2OH}(\textit{trans}) & 16874 & 16649 (7) & 16622 \\
     \ce{CH3CDHOH}(\textit{trans}) & 16836 & 16613 (9) & 16586\\
	\noalign{\smallskip}\hline
	\hline
   
	\end{tabular*}
   
\end{threeparttable}
\end{table}

Table \ref{tab:DMC} shows the DMC ZPEs of ethanol (both isomers) and singly deuterated isotopologues of the \textit{trans} conformer along with semiclassical and harmonic ZPEs. It is seen that the agreement between AS-SCIVR and DMC ZPEs is very good and within method uncertainties (for SC methods uncertainty is typically within 20-30 cm$^{-1}$). Relative to the electronic global minimum, i.e. the bottom of the \textit{trans} conformer well, the DMC ZPEs of \textit{trans} and \textit{gauche} isomers are 17321 $\pm$ 9 cm$^{-1}$ and 17321 $\pm$ 6 cm$^{-1}$, respectively, whereas the corresponding SC ones are 17298 cm$^{-1}$ and 17317 cm$^{-1}$, and the harmonic ZPEs are 17568 cm$^{-1}$ and 17621 cm$^{-1}$. The harmonic ZPEs of the \textit{trans} and \textit{gauche} overestimate the true ZPE values by about 250-300 cm$^{-1}$ revealing a substantial level of anharmonicity. We note that in the DMC calculations very few ``holes'' are detected and in just a couple of trajectories. The total number of ``holes'' detected is 44, which is negligible compared to the total number of configurations (of the order of 10$^{11}$) sampled during the DMC trajectory calculations. This demonstrates that our PES can be in practice considered ``hole-free''. A further certification of this is given by the AS-SCIVR simulations, which are successfully run at energies close to the ZPE one. During DMC propagation, when a random walker encounters a ``hole'' (and thus it enters a region of large potential energy), we kill that walker and let the trajectory continue to propagate. This procedure follows our unbiased DMC algorithm. 

We believe this is the first time the quantum anharmonic ZPE of ethanol is reported at CCSD(T) level of theory. At this point a comparison of our values to the experimentally-derived ones is very insightful. Since the experiment has the ZPE in it, we compare our DMC and SC results with 41 cm$^{-1}$, which is the experimental energy difference value we already anticipated in the Introduction. The bare electronic energy difference on the PES is 38 cm$^{-1}$ with the \textit{trans} conformer being the lower energy one. SC calculations estimate an energy difference of 19 cm$^{-1}$ still in favor of the \textit{trans} conformer, while DMC results have the two conformers basically degenerate. These values suggest an energy gap narrower than the experimentally-derived one, with SC and DMC results in agreement within uncertainty.

\begin{figure}[htbp!]
    \includegraphics[width=0.6\columnwidth]{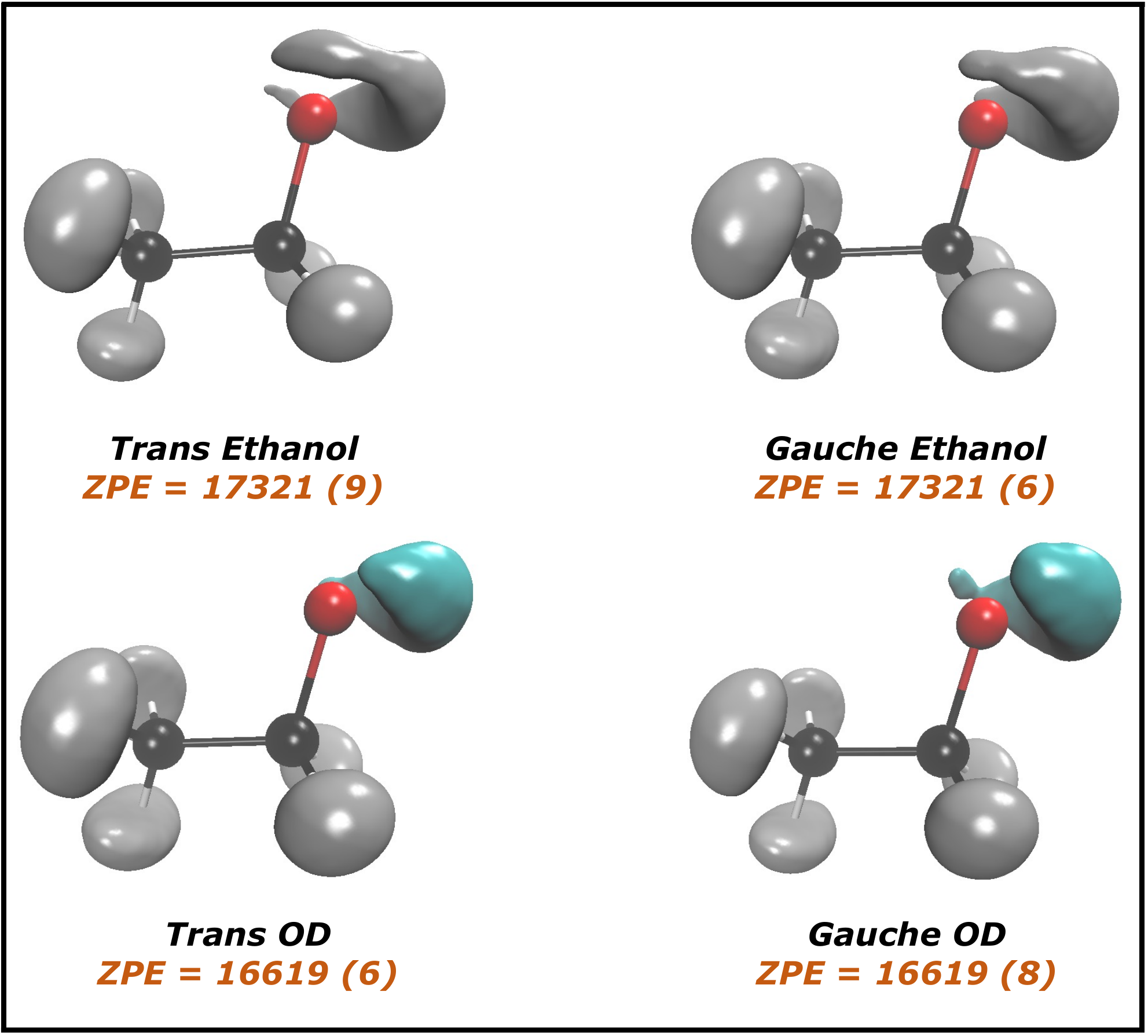}
    \caption{Vibrational ground-state wavefunction. The two upper panels represent the $trans$ and $gauche$-ethanol and the two lower panels represent the $trans$-\ce{CH3CH2OD} and $gauche$-\ce{CH3CH2OD}. The hydrogen atom attached to the oxygen atom has been removed to help the eye. ZPEs values are reported with uncertainties in parentheses.}    
        \label{fig:DMC_wave}
\end{figure}

\begin{figure}[htbp!]
    \includegraphics[width=0.6\columnwidth]{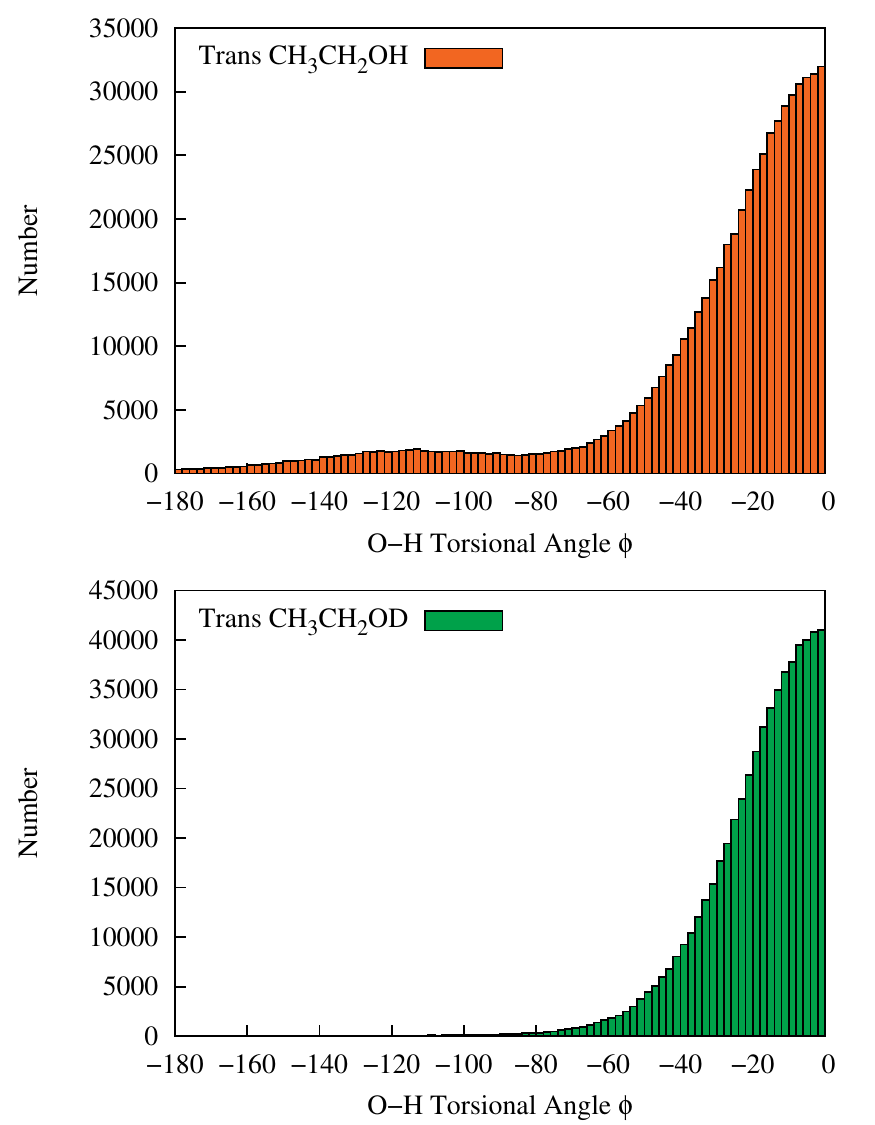}
    \caption{Distribution of C1-C2-O-H torsional angle ($\phi$) from the DMC walkers. The upper panel represents the $trans$-\ce{CH3CH2OH} and the lower panel represents the $trans$-\ce{CH3CH2OD}.} 
        \label{fig:DMC_dist}
\end{figure}

The DMC vibrational ground-state wavefunctions for hydrogens for both \textit{trans} and \textit{gauche} conformers are shown in Figure \ref{fig:DMC_wave}. The DMC results clearly show that the ground-state wavefunction has a \textit{trans} fingerprint even when starting from the \textit{gauche} conformer. On the other hand, the ground-state wavefunction is partly delocalized at the \textit{gauche} geometry. This conclusion is corroborated by the top panel of Figure \ref{fig:DMC_dist}, which shows the distribution of walkers at the end of DMC trajectories (15 DMC trajectories are computed, so total number of walkers are roughly 15$\times$40 000 = 600 000.) started from the \textit{trans} configuration relative to the C1-C2-O-H torsional angle. The \textit{gauche} geometry is found at the  torsional angle of $\pm$120 degrees.  

We also present the vibrational ground-state wavefunction from DMC calculations for the OD motion in $trans$-\ce{CH3CH2OD}, i.e. one of the singly deuterated isotopologues of \textit{trans} ethanol in Figure \ref{fig:DMC_wave}. The ZPEs for the deuterated isotopologue is still equivalent with very similar wavefunctions. In the case of deuteration, the bottom panel of Figure \ref{fig:DMC_dist} shows that the torsional angle distribution is more centered at the \textit{trans} geometry and only very few walkers are found at \textit{gauche} geometry. This shows that, on the one hand, quantum delocalization is somewhat quenched by the deuteration, while, on the other hand, starting from the deuterated \textit{gauche} conformer still leads to the deuterated \textit{trans} one.

The wave function of the OD motion still looks delocalized, but an interesting effect of deuteration on the dynamics of ethanol can be pointed out by examining AS-SCIVR calculations. In fact, as anticipated, a certain rate of AS-SCIVR trajectories are numerically unstable and discarded according to a threshold parameter, as defined in the Theory and Computational Details section. The rejection rate we find is about 55$\%$ for both the \textit{trans} and \textit{gauche} conformers and also for the methyl-deuterated isotopologues. Conversely, for \ce{CH3CH2OD} the rejection rate decreases to about 20$\%$ and 38$\%$ for the \textit{trans} and \textit{gauche} conformer, respectively. This somehow strengthens DMC calculations by providing evidence of a more vibrationally-localized motion for \ce{OD} with respect to \ce{OH} and a clue of a reduced influence of the ``leak'' effect.    

\begin{figure}[htbp!]
    \includegraphics[width=0.8\columnwidth]{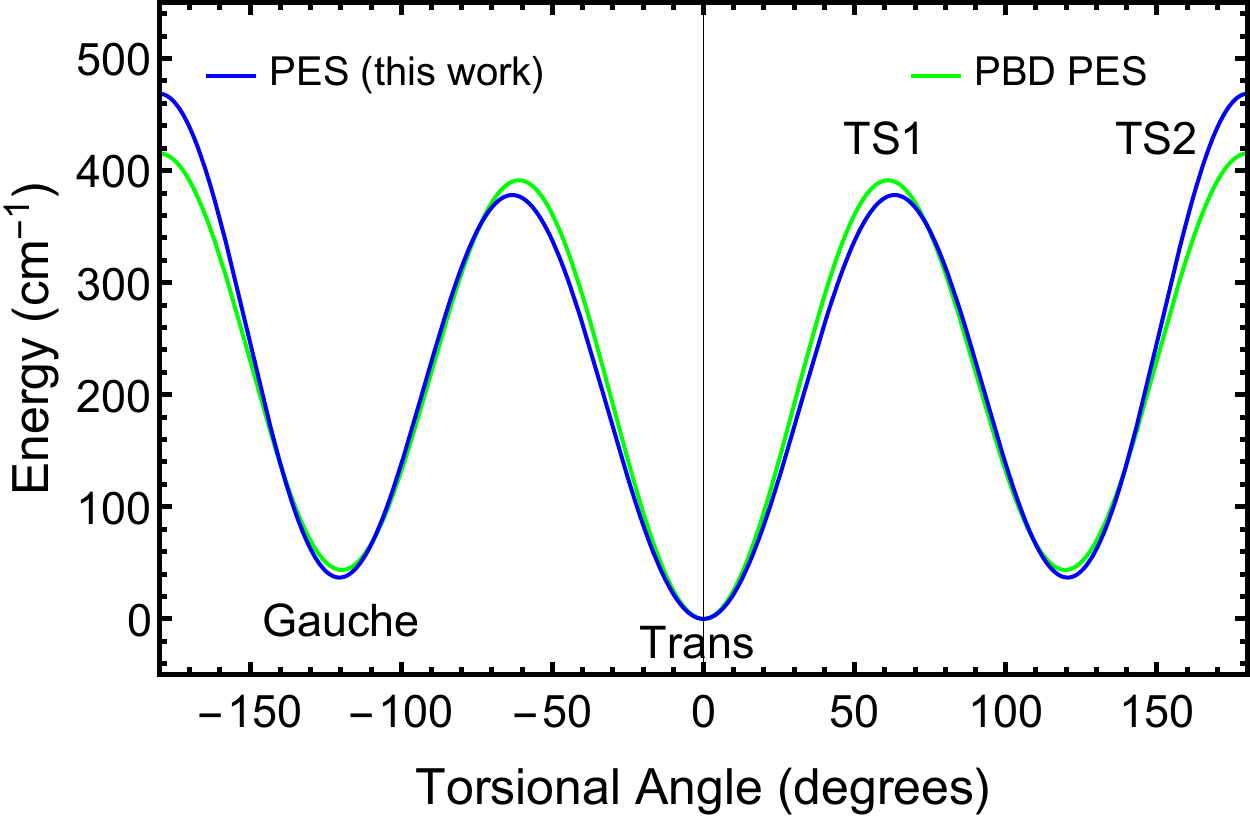}
    \caption{Comparison of C1-C2-O-H torsional potentials from this work (blue) and from PBD\cite{Pearson2008} (green).} 
        \label{fig:PESComparison}
\end{figure}

Then, it is interesting to compare the 1-D O-H torsional potential determined from our full-dimensional PES with the model used by Pearson, Brauer, and Drouin (PBD).\cite{Pearson2008}  As shown in Figure \ref{fig:PESComparison} the two are very similar.  The relative potential energies of the \textit{gauche}  state and TS1 with respect to the \textit{trans} state are nearly the same, while TS2 is somewhat higher in energy for the 1-D potential from our PES as compared to  PBD. Recall that the 1-D OH torsional is not fully relaxed, so some minor adjustments to it might be anticipated.  

\begin{figure}[htbp!]
    \includegraphics[width=0.8\columnwidth]{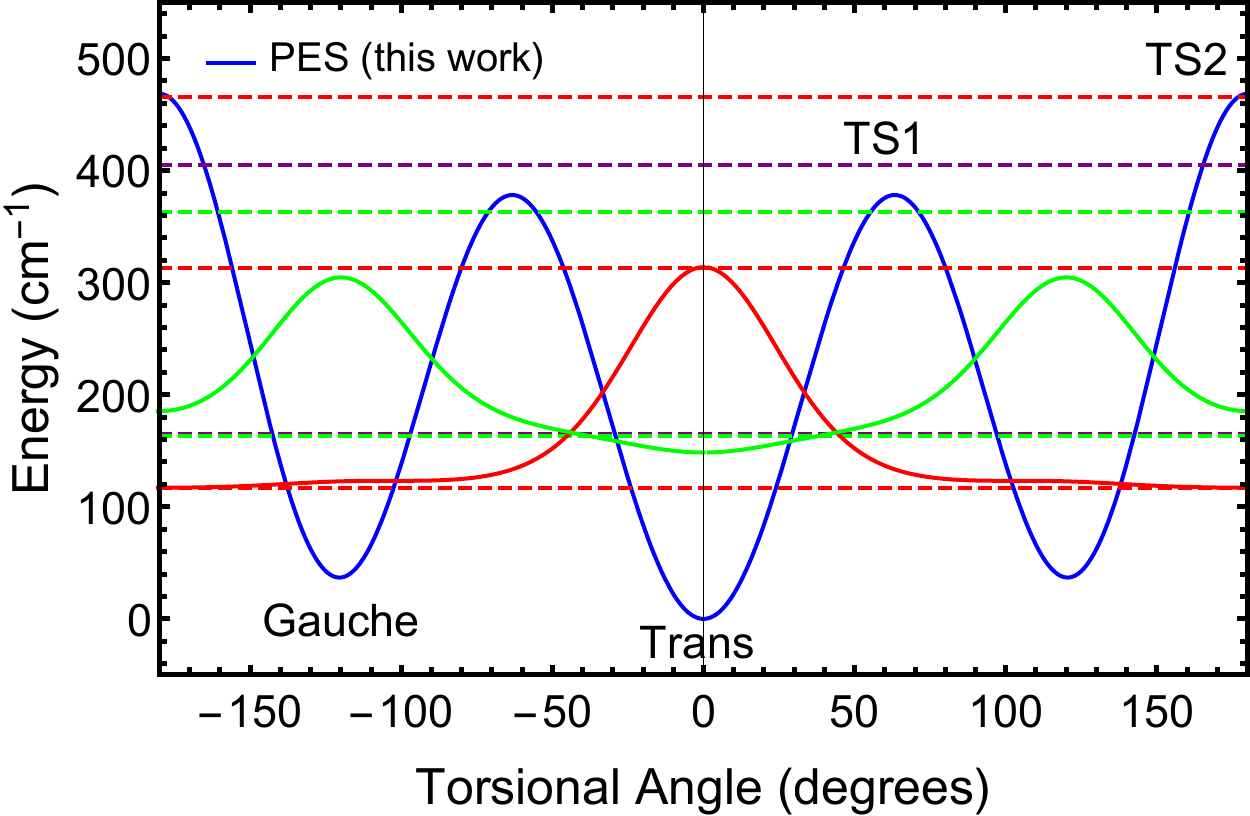}
    \caption{DVR results for energies and wavefunctions based on the 1-D C1-C2-O-H torsional potential from this work. The solid blue curve gives the potential, while the dotted lines give the first seven energy levels (there are two levels at 163.1 and 165.3 cm$^{-1}$).  The solid red and green lines give the wavefunctions corresponding to the two lowest torsional energy levels.} 
        \label{fig:OurPES}
\end{figure}

Having an 1-D model available is always advantageous because one can easily compute the energy levels and the corresponding wavefunctions. Our preferred method for doing so is by using the Discrete Variable Representation (DVR) techniques described in ref. \citenum{dvr1992}.  For the problem at hand, we use the azimuthal (0 to 2 $\pi$ interval, periodic) variant.  There is really only one adjustable parameter, the moment of inertia of the rotor.  An estimate for this might be $\mu_{O-H} \times r_{OH}^2$, where $\mu_{O-H}$ is the reduced mass of the OH in atomic units, and $r_{OH}$ is the equilibrium distance of the O-H bond in bohr. For ethanol, this is about $3.2/(N_{AV} * m_e)$, where $N_{AV}$ is Avogadro's number and $m_e$ is the mass of the electron.  We reduced this numerical value from 3.2 to 2.7 so that, when applied to the PBD model torsional potential, we obtained agreement with their energy differences.   With this moment of inertia then applied to our own PES, we obtained the energy levels and wavefunctions shown in Fig. \ref{fig:OurPES}, where the wavefunctions for only the first two levels are shown.  It is interesting to note that there is substantial wave function amplitude for the \textit{gauche} state at the geometry of the \textit{trans} state and for the \textit{trans} state at the geometry of the \textit{gauche} state, an observation that was shown for the \textit{trans} state also in the DMC results of Figure \ref{fig:DMC_dist} based on the full-dimensional PES.  In fact, the DMC \textit{trans} wavefunction from Figure \ref{fig:DMC_dist} and the wavefunction from Figure \ref{fig:OurPES} are nearly identical, as shown in the Supporting Information in Figure S-5.  

\begin{figure}[htbp!]
    \includegraphics[width=0.8\columnwidth]{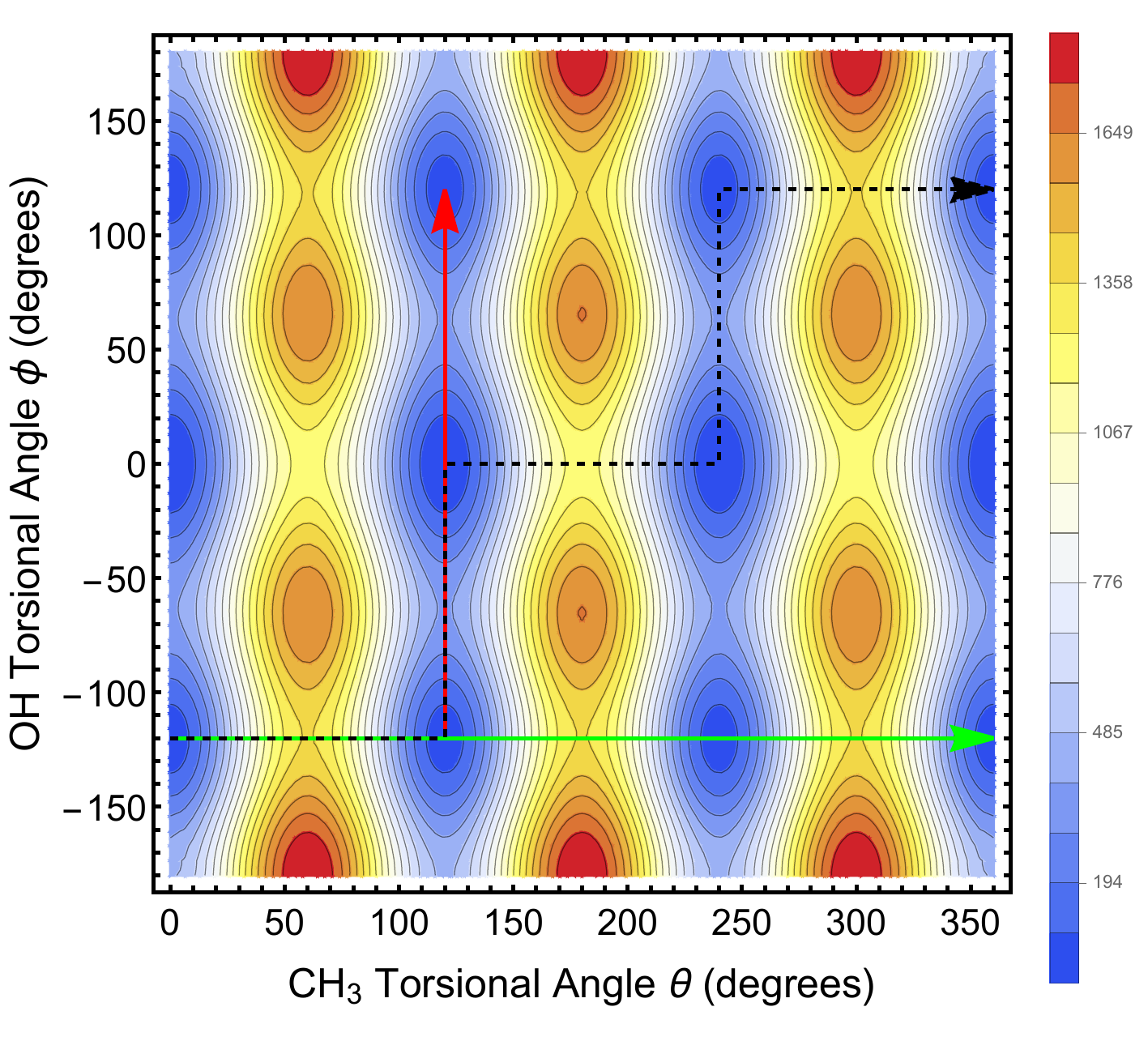}
    \caption{Two-dimensional contour plot of the potential energy as a function of OH torsional angle and \ce{CH3} torsional angle. The color scale gives the potential in cm$^{-1}$.  If the \ce{CH3} torsional angle is constant, for example at 0\textdegree, 120\textdegree, 240\textdegree, or 360\textdegree, the lowest energy path for the OH rotational motion is in the vertical direction, for example along the red arrow.  If the OH torsional angle is constant, for example at -120\textdegree, 0\textdegree, or 120\textdegree, then the lowest energy path for \ce{CH3} rotational motion is in the horizontal direction, for example along the green arrow.  If both the OH and \ce{CH3} are rotating, instead of moving in a straight line, say from \{$\theta,\phi$\}=\{0\textdegree, -120\textdegree\} to \{360\textdegree, 120\textdegree\}, the lowest energy path is to move along the saw-tooth arrow, which describes a geared motion in which the horizontal and vertical displacements take place along the minimum energy paths.}
        \label{fig:gear}
\end{figure}

Of course, a 1-D potential tells only a small part of the story. Two cuts of the 1-D \ce{CH3} torsional potential have previously been shown in Figure \ref{fig:torsional}. When we combine these cuts with two others (taken at the OH torsional angles corresponding to TS1 and TS2, see Figure S-2 in the Supporting Information file) as well as with the OH torsional potential of Figures \ref{fig:PESComparison} and \ref{fig:OurPES}, we can obtain a reasonable fit for a 2-D potential of the combined motions of the OH and the \ce{CH3}, as shown in Figure \ref{fig:gear}. As described in the caption, when both OH and \ce{CH3} are rotating, the minimum energy path for moving, for example, from the well at $\{\theta,\phi\} = \{0^\circ, -120^\circ\}$ to $\{360^\circ,120^\circ\}$ is not at all straight, but rather follows the dashed black saw-tooth path reflecting the geared motion of the two rotors. As anticipated in the Introduction, this geared motion in ethanol has been suggested previously by Quade and colleagues from analysis of microwave spectra, but, to our knowledge, it has not previously been shown via a full-dimensional PES. The functional form we used to fit these potential cuts and then used for Figure \ref{fig:gear} is given in the SI, along with Fig. S-3 showing the DVR results for the \ce{CH3} torsional potential, and Figure S-4, which shows the OH torsional potential for $\theta = 0$ and $\theta = 60$ degrees.

The functional form of the 2-D torsional motions just mentioned also made it possible to perform a 2D DVR calculation of the combined energy levels and wavefunctions.  Moments of inertia of $2.7/(M_{AV} m_e)$ for the OH rotor and $10.5/(M_{AV} m_e)$ for the \ce{CH3} rotor gave the best agreement with the experimental data summarized in PBD.\cite{Pearson2008} The results are shown in Table \ref{tab:PES}, where the first column gives the observed transitions, the second column gives our transition estimates based on the 2-D model (which was fit to five cuts through the full dimensional PES, see SI) and the third column gives the DVR results if instead of the full model potential, we use a separable potential having no cross terms between functions of $\theta$ and $\phi$.  The agreement is good, though certainly not perfect.  It should be noted, however, that the 2-D potential is based on unrelaxed cuts and on a fit to 5 cuts of the potential; other cuts could modify the 2-D fit to the full-dimensional surface.  There could be adjustments due to either effect. Nonetheless, it is remarkable that the \textit{ab initio} surface is in such reasonable agreement with experiment. Following our calculations, we found that Zheng et al. had recommended moments of inertia for the two rotors based on their electronic structure calculations and the resulting low-lying energy levels.  Their results, converted to atomic units, are $2.66/(M_{AV} m_e)$ for the OH rotor and $9.32/(M_{AV} m_e)$ for the methyl rotor, very close to the values we found to be in best agreement with the experimental results of PBD.

\begin{table}[htbp!]
\centering
\caption{Comparison of experimental energy levels\cite{Pearson2008}  relative to the lowest level, our 2-D  DVR calculations, and our 2-D DVR calculations omitting cross terms in the 2D torsional potential. All energies are in cm$^{-1}$.}
\label{tab:exptcompare}
	\begin{tabular*}{1\columnwidth}{@{\extracolsep{\fill}}cccccc}
	\hline
	\hline\noalign{\smallskip}
	 Level & v$_{OH}$ & v$_{CH3}$ &Experiment & full 2-D potential & Omitting cross terms  \\
 \noalign{\smallskip}
	\noalign{\smallskip}\hline\noalign{\smallskip}
 \text{e$_1$} & 0 & 0 & 0 & 0 & 0 \\
 \text{e$_1$} & 0 & 0 & 39.5 & 52.3 & 46.7 \\
 \text{o$_1$} & 0 & 0 & 42.8 & 54.4 & 48.9 \\
  \noalign{\smallskip}
 \text{o$_2$} & 1 & 0 & 202.6 & 198.2 & 196.9 \\
 \text{e$_2$} & 1 & 0 & 238.6 & 236. & 235.1 \\
 \text{o$_3$} & 1 & 0 & 285.9 & 293.6 & 289. \\
  \noalign{\smallskip}
 \text{e$_0$} & 0 & 1 & 244.4 & 251.8 & 246.5 \\
 \text{e$_1$} & 0 & 1 & ? & \text{299.4(?)} & \text{281.9(?)} \\
 \text{o$_1$} & 0 & 1 & ? & \text{301.4(?)} & \text{284.9(?)} \\
 \noalign{\smallskip}
 \text{e$_0$} & 0 & 2 & 475.5 & 
$\begin{array}{c}
 472.1 \\
 477.7 \\
\end{array}$
 & 
$\begin{array}{c}
 468. \\
 473. \\
\end{array}$
 \\
 \noalign{\smallskip}
 \text{e$_1$} & 0 & 2 & 529.49 & 532. & 504.4 \\
 \text{o$_1$} & 0 & 2 & 532.8 & 533.8 & 524.4 \\
    \noalign{\smallskip}\hline
	\hline
	\label{tab:PES}
	\end{tabular*}
\end{table}

\section{Summary and conclusions}
We presented a new potential energy surface for ethanol at the CCSD(T) level of theory.  This was achieved by a $\Delta$-ML method applied to a recent B3LYP-based PES that we previously reported. The new PES was validated for torsional barriers and harmonic frequencies against direct CCSD(T) calculations for the $trans$ and $gauche$ conformers and their isomerization TSs. Diffusion Monte Carlo and semiclassical calculations were reported for the zero-point energies of \ce{CH3CH2OH} and several singly deuterated isotopologues. DMC wavefunctions have also been presented.

Our main goal was to investigate the energetics of ethanol which was known to be characterized by two conformers very close in energy. To achieve this goal we needed a way to perform high-level quantum stochastic and dynamical simulations. Therefore our first effort was to construct a ``gold-standard'' PES of ethanol suitable for quantum calculations that require sampling of the high energy region of the phase space. This is a real need for accurate quantum simulations and not just an exotic requirement. The DMC and SC applications reported demonstrate that not only we achieved our goal, but that the PES is robust for application of methods spanning a large portion of the configuration space. 

Our quantum results provided us with a breakthrough in the chemistry of ethanol since we found that the ground state is of \textit{trans} type with a leak to the \textit{gauche}  conformer. Indeed, DMC ZPE evaluations return the same value starting from both \textit{trans} and \textit{gauche} geometries. A semiclassical estimate of the first excited state starting from the \textit{gauche} conformer provides a reduced energy difference with respect to the energy gap between conformers found by electronic structure calculations. This is also at odds with harmonic estimates, which anticipate an increased gap. In our view, the ``leak'' effect and the reduced energy difference eventually explain experimental discrepancies in ethanol investigations and the difficulty to isolate the two conformers even at low temperatures. 

We also notice that this result points to a striking resemblance with glycine as discussed in one of our previous works.\cite{conte_glycine20} We found that the 8 identified isomers of glycine reduced to 4 couples of conformers once zero-point energy and nuclear dynamics effects were taken into account. However, the impact of this finding was minor compared to the one for ethanol because the three main and experimentally investigated conformers of glycine were still energetically well separated. We think these results, and especially those presented for ethanol, provide a new insight on the chemistry of small organic molecules demonstrating the need to take nuclear quantum effects into account.

We employed the new potential to study the motions of the \ce{-CH3} and \ce{-OH} rotors at the quantum mechanical level. DMC and DVR results are in very good agreement and the computed DVR wavefunctions confirm the presence of the ``leak'' effect. Furthermore, the previously suggested geared motion of the rotors is confirmed by our calculations, and the 2-D model of the torsions based on cuts through the full-dimensional potential provides reasonable energy levels when compared to experiment.

Finally, and as a perspective, we notice that semiclassical calculations are able to evaluate the energy of vibrationally excited states and therefore, given the high fidelity of the PES, they will be employed, together with MULTIMODE calculations, in a future work for determining ethanol fundamental frequencies of vibration. 

\section{Supporting information available}
The supporting information file contains a plot comparing correction, $\Delta{V_{CC-LL}}$ and DFT energies, the harmonic frequencies for the \textit{trans}-\textit{gauche} isomerization saddle points, functional form of the 2-D \ce{CH3} and OH torsional potential, a plot showing the results of a DVR calculation for the \ce{CH3} torsional potential, a plot of the torsional potential of the methyl rotor for TS1 and TS2, the OH torsional potential for $\theta = 0$ and $\theta = 60$ degrees, comparison of the ground state OH torsional wavefunction between DVR and DMC calculations and a schematic of the OH torsional path. 

\section{Acknowledgments}
JMB thanks the ARO, DURIP grant (W911NF-14-1-0471), for funding a computer cluster where most of the calculations were performed and current financial support from NASA (80NSSC20K0360). QY thanks Professor Sharon Hammes-Schiffer and National Science Foundation (Grant No. CHE-1954348) for support. RC thanks Universit\`a degli Studi di Milano (``PSR, Azione A Linea 2 - Fondi Giovani Ricercatori'') for support.
\clearpage

\bibliography{refs}
\clearpage

\section{Supporting Information: Quantum calculations on a new CCSD(T) machine-learned PES reveal the leaky nature of \textit{trans} and \textit{gauche} gas-phase ethanol conformers}

\subsection{Comparison between correction and DFT energies}
\begin{figure}[htbp!]
    \includegraphics[width=0.6\columnwidth]{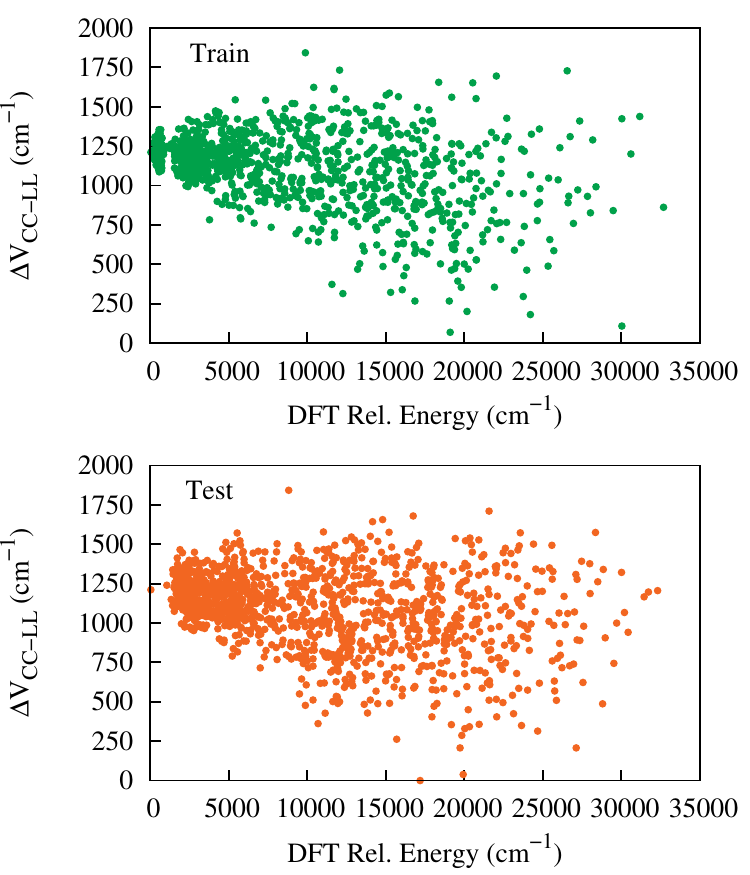}
    \caption{Plot of $\Delta{V_{CC-LL}}$ (relative to the reference value i.e. -35 732 cm$^{-1}$) vs DFT energy relative to the \ce{CH3CH2OH} minimum value with the indicated number of training data sets.}
        \label{fig:Del_plot}
\end{figure}
\newpage
\subsection{Harmonic frequencies:}

   


\subsubsection{Isomerization TSs (Eclipsed and Syn)}

\begin{table}[htbp!]
\centering
\caption{Comparison of harmonic frequencies (in cm$^{-1}$) between $V_{LL{\rightarrow}CC}$ PES and the corresponding \textit{ab initio} (CCSD(T)-F12a/aug-cc-pVDZ) ones of both \textit{eclipsed} and \textit{syn} TSs of Ethanol.}
\label{tab:PES_Freq_TS}

	\begin{tabular*}{0.9\columnwidth}{@{\extracolsep{\fill}}rrrrrrr}
	\hline
	\hline\noalign{\smallskip}
	& \multicolumn{3}{c}{eclipsed} & \multicolumn{3}{c}{syn} \\
	\noalign{\smallskip} \cline{2-4} \cline{5-7} \noalign{\smallskip}
     Mode & \ $\Delta$-ML PES\  & \ \textit{ab initio}\ & \ Diff.\ & \ $\Delta$-ML PES\  & \ \textit{ab initio}\ & \ Diff.\ \\
	\noalign{\smallskip}\hline\noalign{\smallskip}
       1 & 267$i$ & 287$i$ & 20$i$ & 332$i$ & 300$i$ & -32$i$ \\
       2 &  261   &  256   &  -5   &  270   &  271   &   1 \\
       3 &  420   &  416   &  -4   &  411   &  414   &   3 \\
       4 &  800   &  797   &  -3   &  812   &  807   &  -5 \\
       5 &  899   &  899   &   0   &  892   &  892   &   0 \\
       6 & 1058   & 1064   &   6   & 1057   & 1061   &   4 \\
       7 & 1106   & 1106   &   0   & 1105   & 1109   &   4 \\
       8 & 1133   & 1132   &  -1   & 1186   & 1187   &   1 \\
       9 & 1285   & 1285   &   0   & 1307   & 1298   &  -9 \\
      10 & 1370   & 1358   & -12   & 1308   & 1306   &  -2 \\
      11 & 1399   & 1397   &  -2   & 1406   & 1402   &  -4 \\
      12 & 1428   & 1427   &  -1   & 1446   & 1440   &  -6 \\
      13 & 1485   & 1486   &   1   & 1493   & 1493   &   0 \\
      14 & 1500   & 1598   &  -2   & 1507   & 1502   &  -5 \\
      15 & 1522   & 1520   &  -2   & 1534   & 1539   &   5 \\
      16 & 3020   & 3028   &   8   & 3015   & 3027   &  12 \\
      17 & 3028   & 3034   &   6   & 3027   & 3030   &   3 \\
      18 & 3059   & 3069   &  10   & 3054   & 3061   &   7 \\
      19 & 3112   & 3123   &   1   & 3103   & 3106   &   3 \\
      20 & 3123   & 3124   &   1   & 3109   & 3113   &  -6 \\
      21 & 3896   & 3890   &  -6   & 3872   & 3865   &  -7 \\
    \noalign{\smallskip}\hline
	\hline
	\end{tabular*}

\end{table}

\newpage
\subsection{Functional form for the 2-D \ce{CH3} and OH torsional potential and calculations performed with it.}
The functional form of the 2-D fit to the methyl and OH torsional motions shown in the 2-D contour plot of the main text is presented here. The best values of the variables in Table \ref{tab:PES} were obtained by simultaneously fitting five cuts of the OH and \ce{CH3} torsion calculated from the full-dimensional PES. There were two unknown parameters. These cuts are shown in Figs. 4 and 8 of the main text and in Fig. S2, below. The fits are virtually indistinguishable from the data and produced the values shown in the Table.  

\begin{equation} \label{eq:PES}
\begin{split}
V_{OH}(\phi)&= 
     0.5\sum_{n=1}^4 V_{nOH} (1-\text{Cos}( n \phi)),\\
V_{CH3}(\theta)&=V_{CH3}^{\phi=0}(0.5) (1-\text{Cos}(3 \theta)),\\
\text{Correction}(\phi) &= 1+(\sum_{n=1}^3 V_{nx} (1-\text{Cos}(n \phi))\\
V(\theta,\phi)=V_{CH3}&(\text{Correction}(\phi)) \times 
(0.5) (1-\text{Cos}(3 \theta)) + V_{OH}(\phi)
\end{split}
\end{equation}
\noindent
where the values of the constants are listed in the Table below.
\begin{table}[htbp!]
\centering
\caption{Constants for the two-dimensional potential for the OH and \ce{CH3} torsion in ethanol.}
\label{tab:contour_fit}

	\begin{tabular*}{0.5\columnwidth}{@{\extracolsep{\fill}}cc}
	\hline
	\hline\noalign{\smallskip}
	 Constant in Eq. (1) & Value (cm$^{-1}$)  \\
 \noalign{\smallskip}

	\noalign{\smallskip}\hline\noalign{\smallskip}
       $V_{1x}$ & 0.0653\\
      $V_{2x}$ & 0.000147 \\
      $V_{3x}$ & 0.00827 \\
      $V_{CH3}$ & 1208.4\\
      $V_{1OH}$ & 86.3  \\
      $V_{2OH}$ & -4.37  \\
      $V_{3OH}$ & 381.9 \\
      $V_{4OH}$ &  -32.7 \\
       
    \noalign{\smallskip}\hline
	\hline
	\label{tab:PES}
	\end{tabular*}

\end{table}

The 1-D DVR results for the OH torsional potential have been shown in Fig. 8 of the main text.  As mentioned there, the only adjustable parameter is the moment of inertia for the rotor, which was taken to be $2.7/(N_{AV}m_e)$. A 1-D DVR result for the \ce{CH3} potential is shown in Figure \ref{fig:CH3DVR}. The moment of inertia for the methyl rotor was taken here to be $10.5/(N_{AV}m_e)$.

Given the 2D potential in Eq. (\ref{eq:PES}) and the parameters in Table \ref{tab:PES}, we can predict how the OH torsion will vary as a function of the \ce{CH3} torsional angle $\theta$, as shown in Fig. \ref{fig:phiforthetas}.  Not surprisingly, the barriers and the gauche conformation increase in energy as the methyl rotates so that one CH bond eclipses the OH bond. The figure demonstrates substantial interaction between the methyl and OH torsional motions. 

Finally, we can also perform a 2-D DVR calculation\cite{dvr1992} using the model 2-D potential.  The previously described moments of inertia were adjusted to obtain the best fit.  Results are shown in the Table in the main text.

\begin{figure}[htbp!]
    \includegraphics[width=0.6\columnwidth]{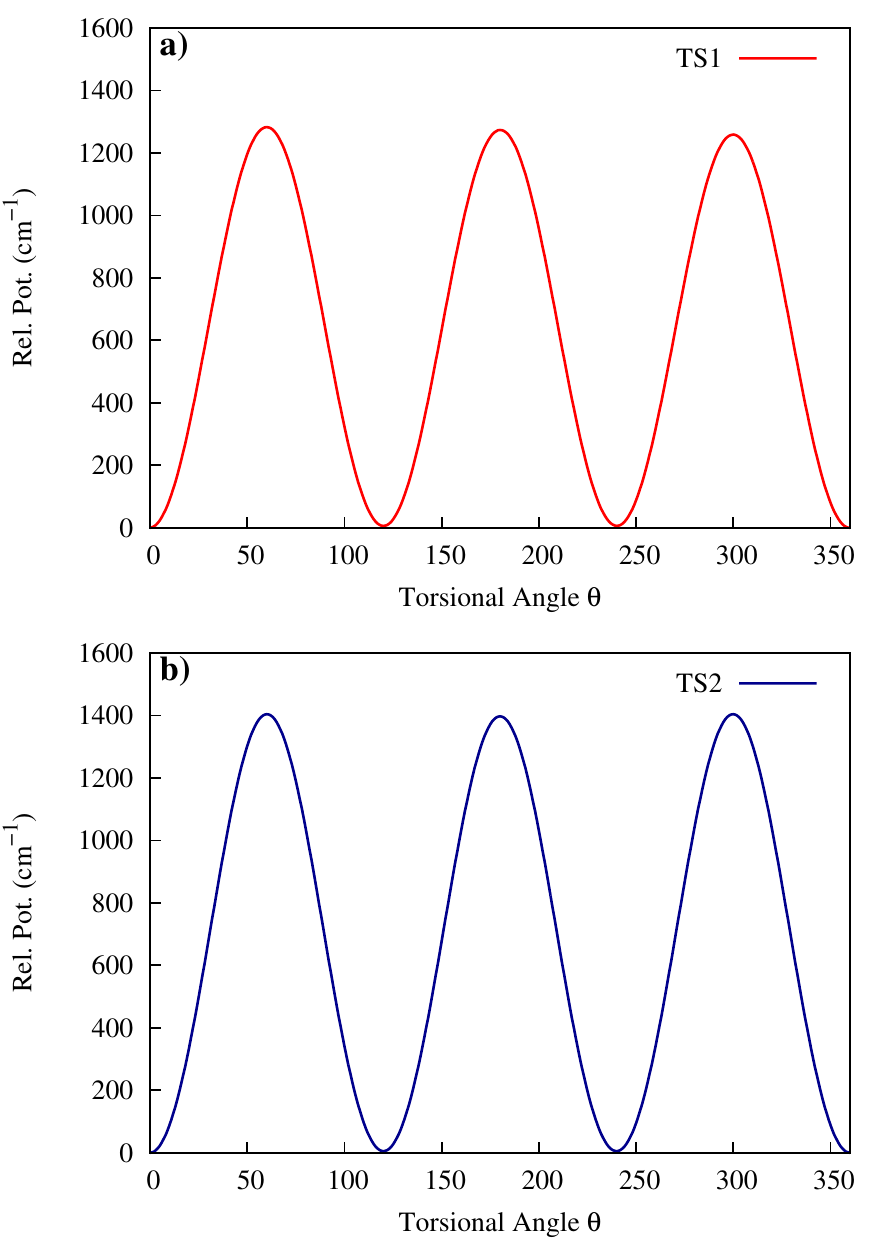}
    \caption{Torsional potential (not fully relaxed) of the methyl rotor of TS1 (a) and TS2 (b) geometry of Ethanol.}
        \label{fig:torsional_TS}
\end{figure}

\begin{figure}[htbp!]
    \includegraphics[width=0.8\columnwidth]{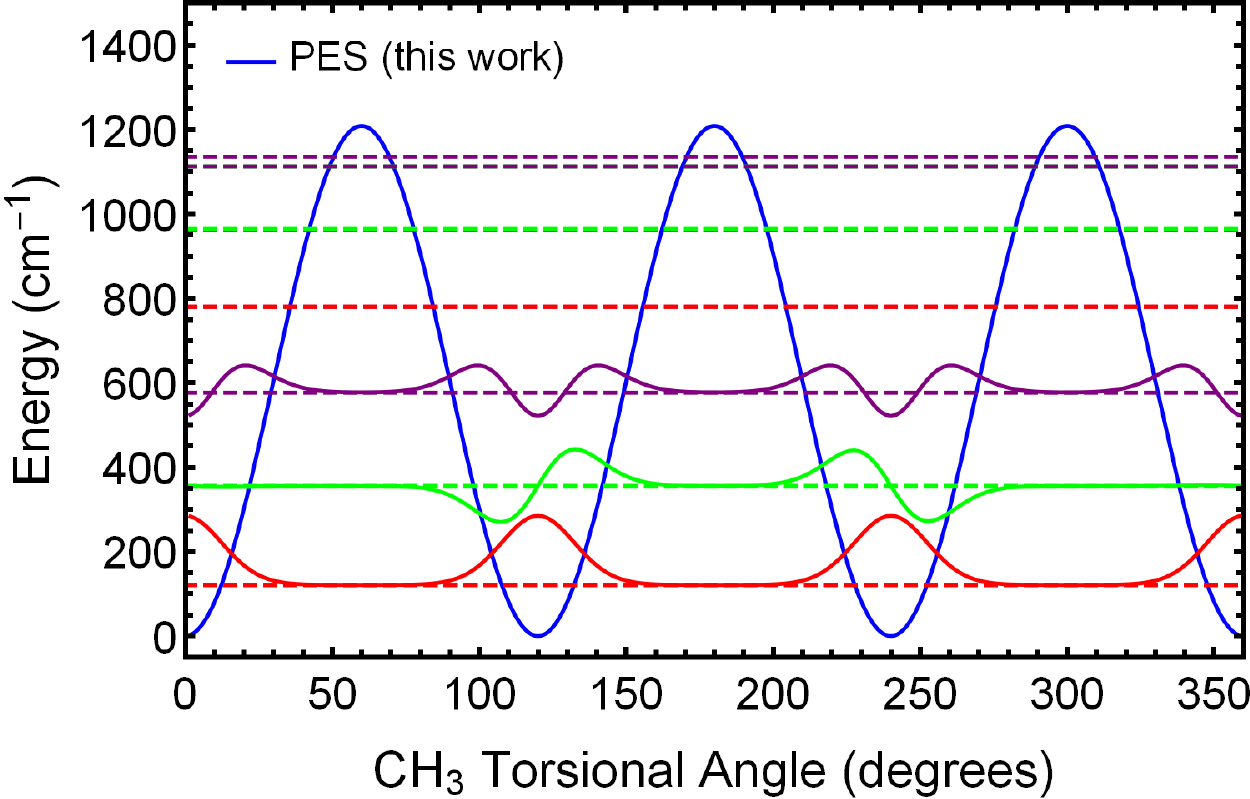}
    \caption{1-D DVR results for the \ce{CH3} torsional potential, whose potential is shown in the blue curve. The energy levels are shown as dotted lines, while the wavefunctions for the lowest three levels are shown as solid red, green, and purple lines. }
        \label{fig:CH3DVR}
\end{figure}

\newpage
\begin{figure}[htbp!]
    \includegraphics[width=0.8\columnwidth]{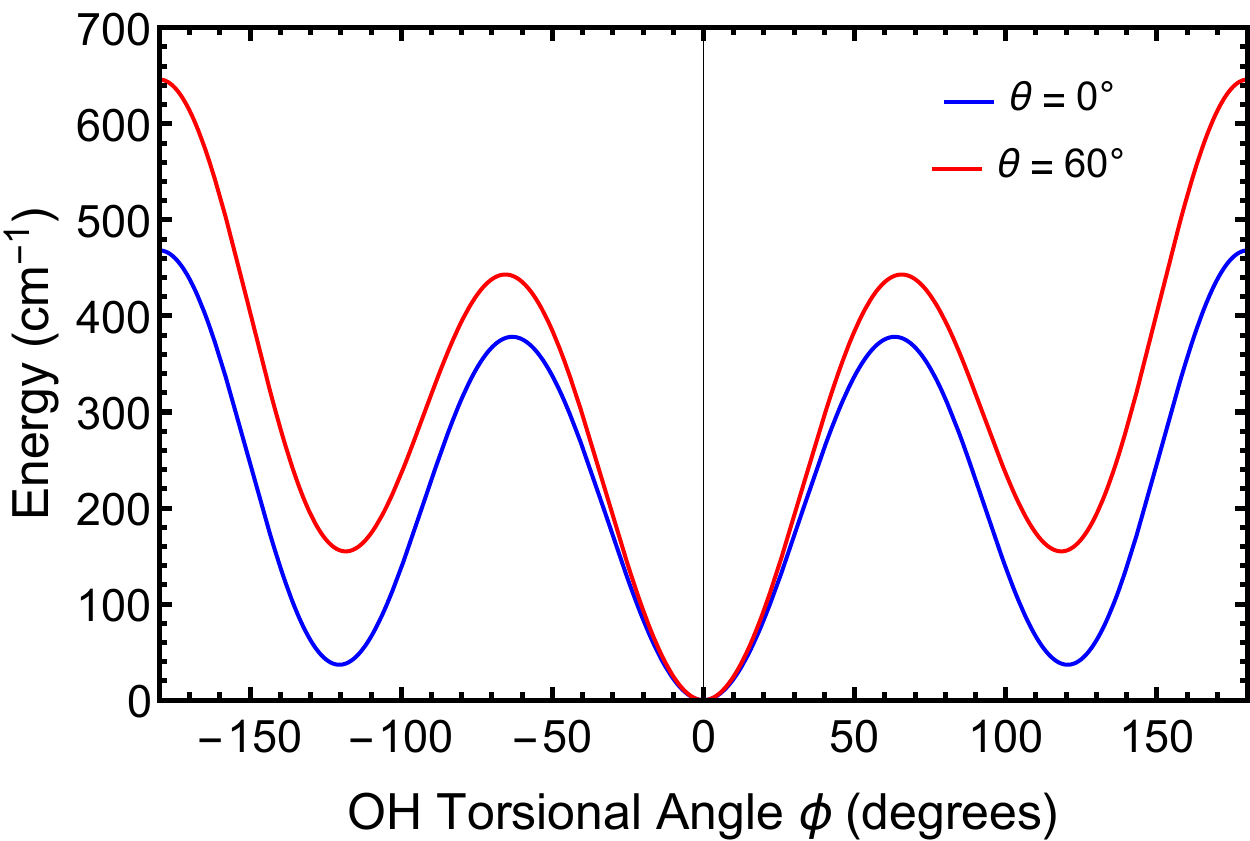}
    \caption{OH torsional potential for $\theta = 0 $ and $\theta = 60$ degrees, normalized to have the same minimum.}
        \label{fig:phiforthetas}
\end{figure}
\newpage
\section{Comparison of DVR and DMC torsional wavefunctions}
\begin{figure}[htbp!]
    \includegraphics[width=0.9\columnwidth]{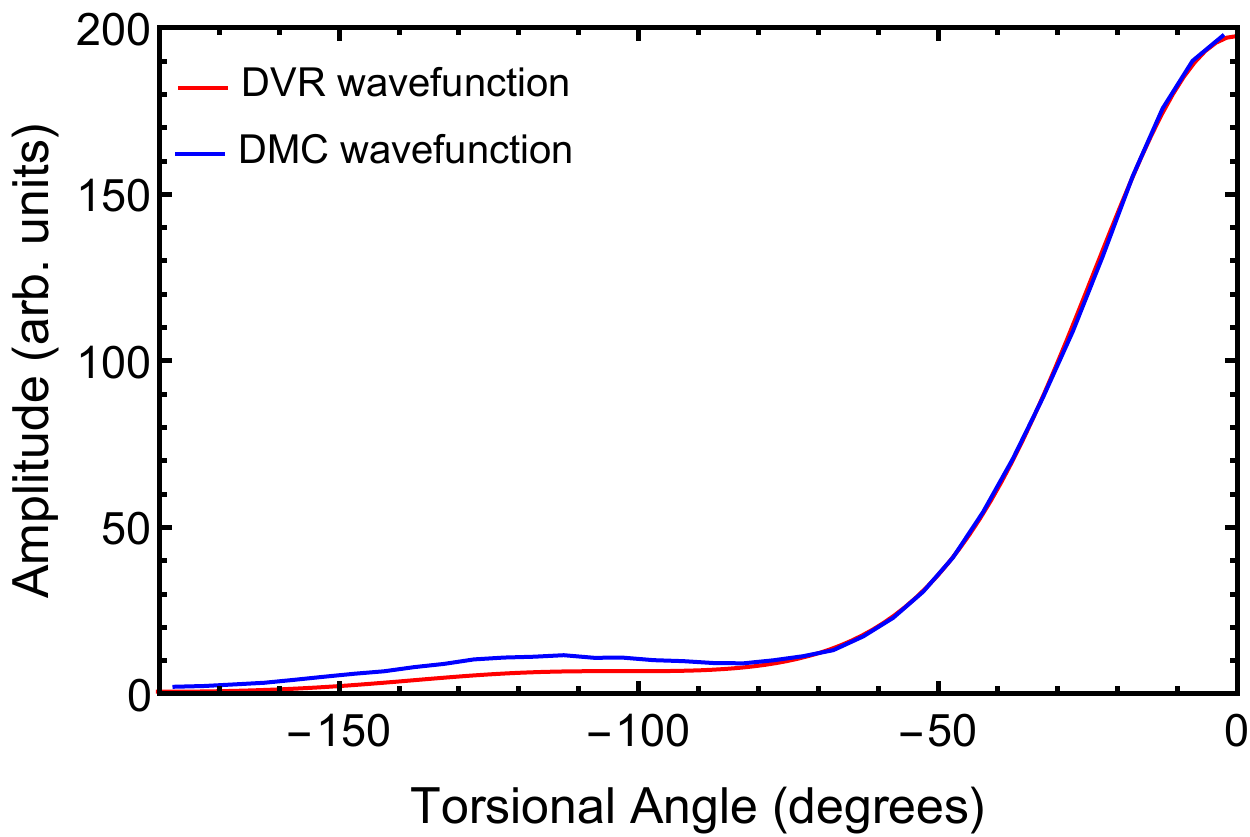}
    \caption{Comparison of the ground state OH torsional wavefunctions as determined from Discrete Variable Representation calculation on a 1-D cut (red) and from Diffusion Monte Carlo calculations on the full-dimensional PES (blue). Note that both wavefunctions have substantial amplitude near $120^\circ$, the geometry of the \textit{gauche} state.}
        \label{fig:wfcompare}
\end{figure}

\begin{figure}[htbp!]
    \includegraphics[width=0.6\columnwidth]{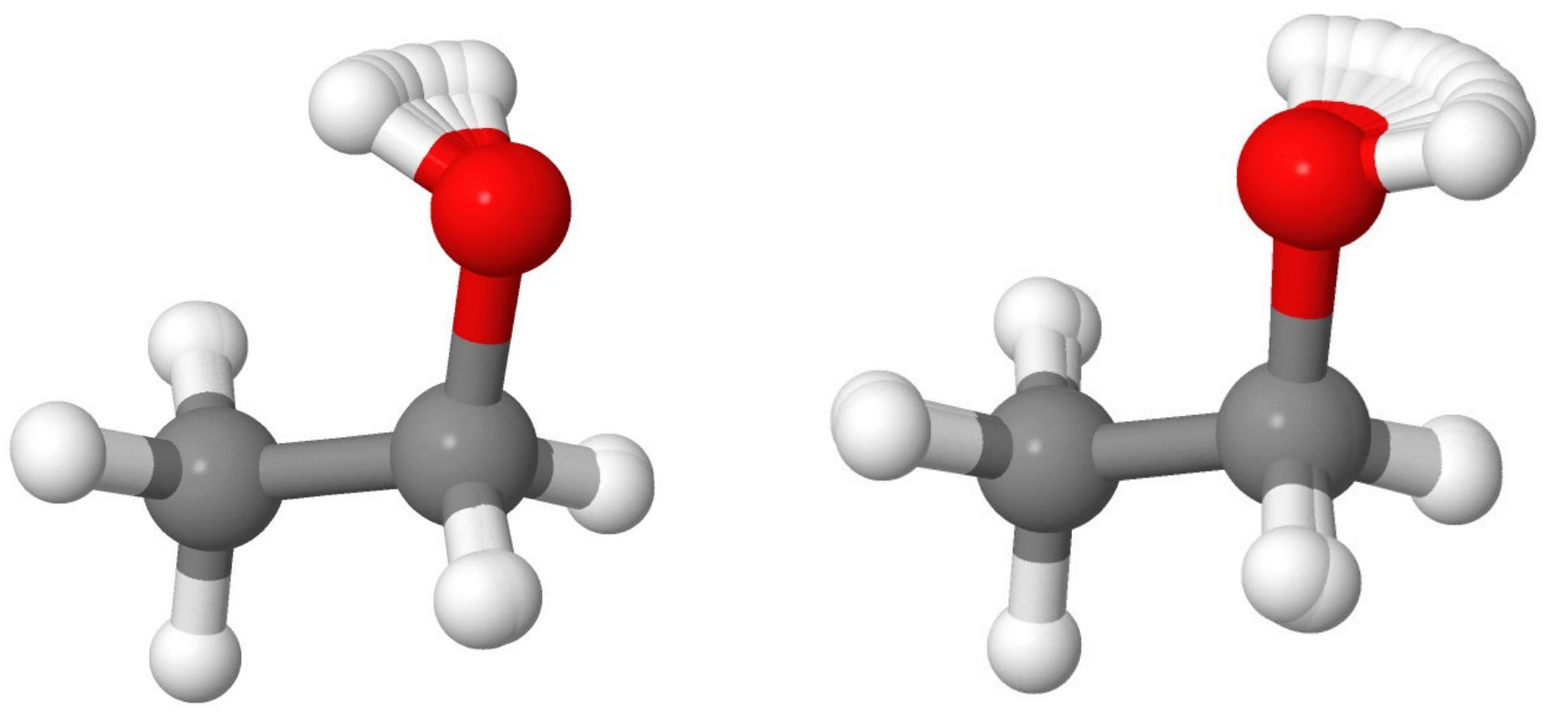}
    \caption{Snapshots of OH torsional path.}
        \label{fig:torsional_path}
\end{figure}

\end{document}